\documentclass[aip,jmp,amsmath,amssymb,reprint,onecolumn]{revtex4-2}
\usepackage{dcolumn}
\usepackage{bm}

\usepackage[utf8]{inputenc}
\usepackage[T1]{fontenc}
\usepackage{mathptmx}
\usepackage{etoolbox}

\usepackage[english]{babel}
\usepackage[utf8]{inputenc}
\usepackage[T1]{fontenc}
\usepackage{amsmath}
\usepackage{hyperref}
\usepackage{graphicx}
\usepackage{amsfonts}
\usepackage{amsthm}
\theoremstyle{remark}

\usepackage{color}
\definecolor{Blue}{rgb}{0.00, 0.00, 1.00}
\definecolor{Red}{rgb}{1.00, 0.00, 0.00}

\newcommand{\nn}{\nonumber}
\newcommand{\be}{\begin{equation}}
\newcommand{\ee}{\end{equation}}
\DeclareMathOperator{\Tr}{Tr}

\newcommand{\moy}[1]{\ensuremath{\left\langle #1 \right\rangle}}

\def\Xint#1{\mathchoice
   {\XXint\displaystyle\textstyle{#1}}%
   {\XXint\textstyle\scriptstyle{#1}}%
   {\XXint\scriptstyle\scriptscriptstyle{#1}}%
   {\XXint\scriptscriptstyle\scriptscriptstyle{#1}}%
   \!\int}
\def\XXint#1#2#3{{\setbox0=\hbox{$#1{#2#3}{\int}$}
     \vcenter{\hbox{$#2#3$}}\kern-.5\wd0}}

\def\dashint{\Xint-}
\newcommand{\Ep}[1]{\mathbb{E}_{\phi}\left[#1\right]}

\makeatletter
\def\@email#1#2{%
 \endgroup
 \patchcmd{\titleblock@produce}
  {\frontmatter@RRAPformat}
  {\frontmatter@RRAPformat{\produce@RRAP{*#1\href{mailto:#2}{#2}}}\frontmatter@RRAPformat}
  {}{}
}%
\makeatother

\begin{document}

\setlength{\abovedisplayskip}{5pt}
\setlength{\belowdisplayskip}{5pt}

\title{Superposition of plane waves in high spatial dimensions:\\ from landscape complexity to the deepest minimum value}

\author{Bertrand Lacroix-A-Chez-Toine}
\affiliation{King's College London, Department of Mathematics, London  WC2R 2LS, United Kingdom}
\email{bertrand.lacroix\_a\_chez\_toine@kcl.ac.uk}
\author{Yan V. Fyodorov}
 \email{yan.fyodorov@kcl.ac.uk}
\affiliation{King's College London, Department of Mathematics, London  WC2R 2LS, United Kingdom}

\begin{abstract}
In this article, we introduce and analyse some statistical properties of a class of models of random landscapes of the form ${\cal H}({\bf x})=\frac{\mu}{2}{\bf x}^2+\sum_{l=1}^M \phi_l({\bf k}_l\cdot {\bf x}), \, \, {\bf x}\in \mathbb{R}^N,\,\, \mu>0 $ where both the functions $\phi_l(z)$ and vectors ${\bf k}_l$ are random. An important example of such landscape
describes superposition of $M$ plane waves with random amplitudes, directions of the wavevectors, and phases, further confined by a parabolic potential of curvature $\mu$. 
 Our main efforts are directed towards analysing  the landscape features in the limit $N\to \infty, M\to \infty$ keeping $\alpha=M/N$ finite. In such a limit we find (i) the rates of asymptotic exponential growth with $N$ of the mean number of all critical points and of local minima known as the annealed complexities and 
(ii) the expression for the mean value of the deepest landscape minimum (the ground-state energy). In particular, for the latter we derive the Parisi-like optimisation functional and 
analyse conditions for the optimiser to reflect various phases for different values of $\mu$ and $\alpha$:  replica-symmetric, one-step and full replica symmetry broken, as well as criteria for continuous, Gardner and random first order transitions between different phases.  
\end{abstract}

\maketitle

\section{Introduction}

High-dimensional random landscapes play a central role in the description of many complex systems, ranging from spin-glasses to non-convex optimisation problems underlying inference and machine learning models, all the way to  modelling biological fitness, see \cite{ros2023high} for a review on this subject and further references. One of the central questions associated to these landscapes is to describe properties of its global minimum, also frequently called in the spin glass context the ground-state energy, and developing methods and/or algorithms to reach the ground-state configuration. Recent years have seen phenomenal progress in obtaining accurate approximate solutions of such problems via machine-learning algorithms, whose use has now become widespread. A detailed description of the arrangement and properties of critical points (maxima, minima and saddles) as well as understanding statistics of the ground state and associated Hessians in the simplest models of random landscapes already provided a deeper understanding of such non-convex optimisation problems and helped to create efficient algorithms, see \cite{AuffingerMontanariSubag23,HuangSellke24} for discussions of recent developments. A large body of literature in theoretical physics as well as in mathematics has been devoted in recent years to studying these questions in the framework of a few simple models, mainly with Gaussian disorder. Obviously, adding to the pool of analytically tractable models is of considerable interest both to get closer to realistic situations and to investigate questions of universality.

In this article, we will analyse both the mean ground-state energy and the problem of counting the number of critical points and local minima for random landscapes ${\cal H}({\bf x})$ of a particular form 
\be
{\cal H}({\bf x})=\frac{\mu}{2}{\bf x}^2+\sum_{l=1}^M \phi_l({\bf k}_l\cdot {\bf x}), \, \, \mu>0\label{H}
\ee
defined on a continuous configuration space ${\bf x}\in \mathbb{R}^N$. Here the disordered part is the sum of $M$ random {\it stationary} processes $\phi_l(z)$ with zero mean $\mathbb{E}\left[\phi_l(z)\right]=0$,  independent and identically distributed (i.i.d.) for different $l$,  whose statistical properties remain invariant with respect to translations in the argument $z\to z+a$ for any real shift $a$.  Two explicit particular examples of such functions $\phi_l(z)$ are described in more detail below, and we frequently concentrate on them in our exposition, but most of the obtained results remain valid for a larger class of functions satisfying the assumption of stationarity. 
\begin{enumerate}
    \item Our first example corresponds to Gaussian-distributed stationary random processes $\phi_l(z)$ with zero mean, all the statistical properties being therefore controlled by the covariance, which is assumed to be of the form 
\be
\Ep{\phi_l(z)\phi_m(z')}=\delta_{lm}R\left(|z-z'|\right)\;,\;\;R(w)=\int_{-\infty}^{\infty} \frac{dq}{2\pi}\tilde R(|q|)\cos(q w)\;.
\ee
Realisations of such random functions are most conveniently generated via introducing its Fourier transform
\be
\phi_l(z)=\int_{-\infty}^{\infty}\frac{dq}{2\pi}\,\left[u_l(q)\cos(q z)+v_l(q)\sin(q z)\right]
\ee
with Gaussian-distributed delta-correlated  Fourier coefficients
\be 
\mathbb{E}\left[u_l(q_1)u_m(q_2)\right]=\mathbb{E}\left[v_l(q_1)v_m(q_2)\right]=2\pi \delta_{lm}\tilde R(|q_1|)\delta(q_1+q_2)\;,\;\;\mathbb{E}\left[u_l(q_1)v_m(q_2)\right]=0\;.
\ee
 One of the simplest representative of this class is the process
\be
\phi_l(z)=u_l\cos(z)+v_l\sin(z)\;,\label{sum_cos}
\ee
where $u_l,v_l$ are i.i.d. standard real normal random variables (with zero mean and unit variance).
In this particular case characterized by the covariance $\mathbb{E}\left[\phi(z)\phi(z')\right]=\cos(z-z')$ some properties of the arising random landscape Eqs.\eqref{H}, notably its expected number of critical points, have been investigated in our earlier paper \cite{lacroix2022superposition}. 
\item Another class of random functions considered in our present paper corresponds to random fields explicitly constructed as
\be
\phi_l(z)=\sum_{n=1}^{\infty}\gamma^{(n)}_l\cos\left(n\left(z+\theta^{(n)}_l\right)\right)\;, \label{sumcos}
\ee
with phases $\theta^{(n)}_l$'s that are i.i.d. random variables for different $n$, uniformly distributed over $[0,2\pi)$, which ensures stationarity of the process. The random amplitudes $\gamma^{(n)}_l$ are independent  of the phases $\theta^{(n)}_l$, independent between different $n$, and can be drawn from any distributions  with a finite value of $\mathbb{E}\left[ \left|\phi_l(z)\right|\right]<\infty$. Every realization of such processes is obviously $2\pi$-periodic. The simplest example of such periodic random functions corresponds to including only the first harmonic ($n=1$) in the above sum 
\be
\phi_l(z)=\gamma_l\cos(z+\theta_l)\;.\label{phi_theta}
\ee
Note that choosing in \eqref{phi_theta} $\gamma_l>0$ with the probability density $p(\gamma_l)=\gamma_le^{-\frac{\gamma^2_l}{2}}$ makes this case equivalent to \eqref{sum_cos}.

 In Eq. \eqref{H} the random functions $\phi_l(z)$  are evaluated at random argument values $z_l={\bf k}_l\cdot {\bf x}$ where ${\bf k}_l$'s are wavevectors taken either as  i.i.d. random vectors uniformly sampled on the sphere $\sqrt{N}{\cal S}_{N-1}$ or, alternatively, as vectors with $N$ i.i.d. normally distributed entries, each with zero mean value and variance $1/N$. 

It is worth noting that the random potential built by superposition of $\phi_l(z)$ described in \eqref{phi_theta}  has an interesting connection to the context of semiclassical chaos.  In the latter framework one may consider a Riemannian manifold ${\cal D}$ with strongly chaotic classical flow, and exploit the  so-called Berry's conjecture\cite{berry1977regular} for the eigenfunctions of the associated quantum Laplacian 
\be
-\Delta \psi_n({\bf x})=E_n\,\psi_n({\bf x})\;,\;\;{\bf x}\in {\cal D}\subset \mathbb{R}^N\;,
\ee
where $\Delta$ is the corresponding Laplace-Beltrami operator together with appropriate boundary conditions. For the case of Euclidean space the latter is the usual Laplacian 
and the eigenfunctions  can be expressed as a sum over plane waves with fixed length of the wave vectors:
\be\label{Berry}
\psi_n({\bf x})=\sum_{l} \gamma_l\cos({\bf k}_{n,l}\cdot {\bf x}+\theta_l)\;,\;\;{\bf k}_{n,l}^2=E_n\;.
\ee
According to Berry such eigenfunctions for high enough eigenvalues $E_n$ should be described in the form Eq. \eqref{Berry} with i.i.d. Gaussian amplitudes $\gamma_l$ and taking the wavevectors ${\bf k}_{n,l}$ independent and randomly equidistributed over the unit sphere of radius $\sqrt{E_n}$. The random part of the potential defined in Eq. \eqref{H} is then identical with a specific realisation of such semi-classical chaotic wave-function $\psi_n({\bf x})$, considering the number of contributing wavevectors to be essentially infinite: $M = \infty$.
In that context the problem of analysing statistics of minima, maxima and saddles of arising wavefunctions and closely related random fields on two-dimensional manifolds $N=2$, as well as  of topology of the associated nodal set
was intensively studied over the years both in mathematical  \cite{cammarota2016distribution,cammarota2017fluctuations,BeliaevWigmanCammarota19,beliaev2020no,Wigman24}  as well as physical (see e.g. \cite{Jain2017nodal}) literature.  
The statistics of the height of the global maximum for such eigenfunctions has been addressed in that context  as well \cite{aurich1999maximum}.

In the present paper we analyse not dissimilar questions employing however a somewhat different limit 
\be
N,M\to \infty\;,\;\;{\rm with}\;\;\alpha=\frac{M}{N}=O(1)\;.\label{hd_lim}
\ee
Although we do not pursue this line explicitly in the paper, we believe some of our results may still be relevant for statistical properties of the chaotic semi-classical wave-functions on high-dimensional manifolds,  after interpreting the parameter $\mu$ as the inverse squared radius of the domain occupied by the wavefunction
and further considering $\alpha \to \infty$. Whether our results for finite $\alpha$ may still have some meaning in the context of semiclassical quantum chaos on high-dimensional manifolds remains an interesting open question.

\end{enumerate} 

To put our research in a proper context let us recall that random potential landscapes of the form Eq. \eqref{H} but with random part replaced with  a Gaussian-distributed function $V({\bf x})$ with zero average $\mathbb{E}\left[V({\bf x})\right]=0$ and covariance of the form
\be
\mathbb{E}\left[V({\bf x}_1)V({\bf x}_2)\right]=N\,F\left(\frac{({\bf x}_1-{\bf x}_2)^2}{2N}\right)\;\label{GRF}
\ee
have been attracting much of attention in the disordered systems' literature since the 90's as a toy model for an elastic manifold of zero internal dimension placed in a random potential. Its dynamical properties \cite{franz1994off,cugliandolo1995dynamics}, thermodynamic properties \cite{mezard1990interfaces,mezard1991replica,mezard1992manifolds,engel1993replica,fyodorov2007classical}, the number of critical points \cite{fyodorov2004complexity,fyodorov2007replica} and Hessian spectrum at the ground state \cite{Hessian_FyoLD_18} have all been studied extensively in the physical literature, predominantly by the heuristic replica approach. The rigorous mathematical studies of the toy model as well as its extensions to elastic manifolds of finite internal dimension have been very recently developed in greater generality as well\cite{BenArousBourgadeMcKenna24,Xu2022hessian,BenArous_Kivimae1,BenArous_Kivimae2}. 
Most importantly, it has been established that this class of random landscapes displays a glass-like {\it ergodicity breaking transition} which translates into many observables of these disordered systems. The perhaps simplest characterisation for this transition is obtained by computing the so-called "annealed complexity" of critical points, defined in the high-dimensional limit $N\to \infty$ as
\be
\Sigma_{\rm tot}(\mu)=\lim_{N\to \infty}\frac{1}{N}\ln \mathbb{E}\left[{\cal N}_{\rm tot}\right]\;,\label{an_comp}
\ee
where ${\cal N}_{\rm tot}$ denotes the total number of stationary points of the random landscape. The {\it topology trivialisation transition}  \cite{fyodorov2004complexity,fyodorov2007replica,fyodorov2014topology,Fyo_2015_rev} 
describes the transition in the number of critical
points from exponential in $N$ to sub-exponential beyond a critical value of the control parameters (in the present model its role is played by the curvature parameter $\mu$).   The true location of the ergodicity breaking/topology trivialization transition should however be obtained from the point of vanishing of the associated quenched complexity, defined instead as $\Xi_{\rm tot}(\mu)=\lim_{N\to \infty}\mathbb{E}\left[\frac{1}{N}\ln {\cal N}_{\rm tot}\right]$  and thus upper-bounded by the annealed complexity. Although in general the quenched and annealed complexities are expected to differ,  they  nevertheless were found to coincide in a few cases of the models of not dissimilar nature, see  \cite{Subag2021concentration,Kivimae2023ground}. Similar annealed complexities can be computed for the number of minima \cite{Fyodorov_Nadal_12}, as well as for critical points of any index conditioned on the height of the landscape, see  \cite{AufBenArusCerny2013,AufBenArous2013}. The latter is especially important as it allowed to relate the annealed complexities computations to the properties of the ground state, the ultimate goal of optimisation problems. 

An alternative characterisation of the ergodicity breaking transition in the landscape paradigm can be obtained by computing the mean (which is simultaneously typical) value of the ground-state energy 
\be
e_0=\lim_{N\to \infty}\frac{1}{N}\min_{{\bf x}}{\cal H}({\bf x})=\lim_{N\to \infty}\frac{1}{N}\mathbb{E}\left[\min_{{\bf x}}{\cal H}({\bf x})\right]\;.\label{GSE}
\ee
This quantity has historically been computed in the physics literature using the replica method for evaluating the associated free energy at finite temperatures as pioneered by Giorgio Parisi  \cite{mezard1987spin} (see 
a modern exposition in the book\cite{parisi_urbani_zamponi2020book}), with alternative rigorous methods  developed by mathematicians in the last two decades \cite{guerra2003broken,talagrand2006parisi,Talagrand_spherical,Panchenko2013,Subag2021groundstate,DominguezMourrat_book_23}. In such a context, the mean free energy is obtained by solving an associated optimisation problem for the so-called Parisi functional, and  the ergodicity breaking transition manifests itself via the {\it replica symmetry breaking} transition, and described by the changes in the nature of the associated optimiser, see e.g. \cite{auffinger2015properties}. In such a framework studying the ground-state energy requires to consider the limit of zero temperature of the optimisation problem. The corresponding studies provided valuable information on the ergodicity broken phase \cite{AufChen2017,ChenSen2017,auffinger2020sk,AufZhou2022}. In this way one was able to characterise the nature of the replica symmetry breaking to be either one-step replica symmetry broken (1RSB), full replica symmetry broken (FRSB), or more exotic patterns \cite{crisantileuzzi2004,crisantileuzzi2006,AufZeng2019} as well as transitions between such patterns. Moreover, the full distribution of overlaps between configurations corresponding to different global minima at the same energy may eventually be computed. Obtaining such detailed information through the investigation of the arrangements of critical points of the landscape would require an in-depth characterisation of the quenched complexity, which is still missing for most models (see however \cite{subag2017complexity,Subag2021concentration, Kivimae2023ground}). {\it A priori} it is therefore not at all evident that vanishing of the annealed complexities indeed signals the ergodicity breaking. Nevertheless, in the toy Gaussian model it was observed to be the case long ago\cite{fyodorov2007replica}. More  recently it was also conjectured to hold for the Gaussian  spherical model \cite{belius2022triviality} and very recently rigorously  proved for a multispecies version of the latter \cite{Huang_Sellke23}.

Finally, it is worth also noting that the landscape complexities for the case of {\it non-random} functions $\phi_l(x)=\phi(x), \, \forall l$ in the model Eq. \eqref{H}, with vector ${\bf x}$ further restricted to a sphere, have been addressed in \cite{maillard2020landscape}. Randomness in the choice of $\phi_l(x)$ makes however our model conceptually very different.

The remaining part of this article is organised as follows. In the following section \ref{sec_main_res}, we summarise our main results both on the annealed complexities and on the mean ground state energy. In section \ref{sec_complexity} we present a computation of our results for the complexities, starting from the Kac-Rice formula. In section \ref{sec_opt_prob}, we employ the replica trick to derive a general functional optimisation problem allowing to express the ground-state energy in the zero-temperature limit. In section \ref{sec_analysis}, we analyse that optimisation problem in different thermodynamic phases and obtain the transition lines. Finally in section \ref{sec_conclu}, we give concluding remarks and mention some future directions. Some technical details are relegated to the appendices.

\section{Summary of the main results} \label{sec_main_res}
 The results of this paper are largely derived at the level of rigour typical for practitioners in the field of Theoretical Physics. Most importantly, the derivation of the optimisation problem for the free energy relies on the use of the replica trick  employing the Parisi scheme of replica symmetry breaking. In the process of computing annealed complexities we rely upon highly plausible but not yet fully rigorously verified strong self-averaging property of certain random matrix determinants, and also employ large deviation techniques in a heuristic form based on functional integrals.   
 Therefore, at the level of mathematical statements our results must be understood as well-grounded conjectures. We hope that they will eventually be amenable to a proof with full mathematical rigour.  

\subsection{Results on the annealed complexities}

The mean total number of critical points $\mathbb{E}\left[{\cal N}_{\rm tot}(\mu)\right]$ and the mean number of minima $\mathbb{E}\left[{\cal N}_{\min}(\mu)\right]$ for random landscapes in the class defined in Eqs. (\ref{H}) can be computed explicitly  for 
any finite $N$ and $M$ using the Kac-Rice approach, see e.g. \cite{adler2007random,azais2009level} for fully rigorous introduction and \cite{maillard2020kac} for a nice informal treatment and further references. They are given, respectively, by
\begin{align}
  \mathbb{E}\left[{\cal N}_{\rm tot}(\mu)\right]&=\frac{1}{\mu^N}\mathbb{E}\left[\left|\det\left(\mu\mathbb{I}+K T K^T\right)\right|\right]\;,\label{n_tot_res}\\
  \mathbb{E}\left[{\cal N}_{\min}(\mu)\right]&=\frac{1}{\mu^N}\mathbb{E}\left[\det\left(\mu\mathbb{I}+K T K^T\right)\Theta\left(\mu\mathbb{I}+K T K^T\right)\right]\;,
\end{align}
where the columns of the rectangular matrix $K$ of size $N\times M$ correspond to the wavevectors $k_l, \, l=1,\ldots,M$'s while the $M\times M$ matrix $T$ is diagonal with entries $T_{lm}=\delta_{lm}\phi_l''(z)$, with properties of random functions $\phi(z)$ discussed after Eq. \eqref{H} . Correspondingly, $T_{ll}=\phi_l''(z)$ are i.i.d. random variables, whose probability density function (PDF) is independent of $z$ and will be denoted as $p_0(t)$. We further set $p_0(t)dt:=dP_0(t)$ and assume, in particular, that the PDF satisfies $\int |t|\,dP_0(t)<\infty$. The function $\Theta(A)$ is an indicator function ensuring that the smallest eigenvalue of the matrix $A$ is positive.

In the high-dimensional limt $N,M\to \infty$ with $\alpha=M/N=O(1)$, the total annealed complexity reads
\begin{align}
 \Sigma_{\rm tot}(\mu,\alpha)&=\lim_{N,M\to \infty}\frac{1}{N}\ln \mathbb{E}\left[{\cal N}_{\rm tot}(\mu)\right]=\int_{\mu}^{\infty}\left(\frac{1}{\nu}+m_{r}(-\nu;\alpha)\right)\,d\nu\,;,
\end{align}
where the two functions $m_r\equiv m_{r}(-\nu;\alpha)$ and $m_i\equiv m_{i}(-\nu;\alpha)$ satisfy the following set of integral equations
\begin{align}
    \frac{m_{r}}{m_{r}^2+m_{i}^2}&=-\nu-\alpha\frac{\displaystyle\int 
    \frac{t(1-t\,m_{r})}{\sqrt{(1-t\, m_{r})^2+t^2 m_{i}^2}}\,dP_0(t)}{\displaystyle\int \sqrt{(1-t\, m_{r})^2+t^2 m_{i}^2}\,dP_0(t)\,}\;,\label{m_r_eq_res}\\
    \frac{m_{i}}{m_{r}^2+ m_{i}^2}&=\alpha m_i\frac{\displaystyle\int \frac{t^2}{\sqrt{(1-t\, m_{r})^2+t^2 m_{i}^2}}\,dP_0(t)}{\displaystyle\int \sqrt{(1-t\, m_{r})^2+t^2 m_{i}^2}\,dP_0(t)}\;.\label{m_i_eq_res}
\end{align}
Note that $m_r$ and $m_i=\pi \rho(-\nu)$ correspond respectively to the real and imaginary parts of the Stieltjes transform $m(-\nu;\alpha)=-\int \frac{\rho(\lambda)\,d\lambda}{\lambda+\nu}$ of the limiting density of eigenvalues $\rho(\lambda)$ of the matrix $K T K^T$ at the spectral point $\lambda=-\nu$. 
The integration in the above formulae goes over the whole real line $t\in \mathbb{R}$.

The annealed complexity of minima is given by a very similar expression:
\begin{align}
 \Sigma_{\min}(\mu,\alpha)&=\lim_{N,M\to \infty}\frac{1}{N}\ln \mathbb{E}\left[{\cal N}_{\min}(\mu)\right]=\int_{\mu}^{\infty}\,\left(\frac{1}{\nu}+m(-\nu;\alpha)\right)\,d\nu\;,
\end{align}
albeit now $m\equiv m(-\nu;\alpha)$ satisfies together  with $t_{\rm m}\equiv t_{\rm m}(\nu;\alpha)$ a different set of integral equations:
\begin{align}
    \frac{1}{m}&=-\nu-\alpha\frac{\displaystyle\int_{-t_m}^{\infty}\,t\,dP_0(t)}{\displaystyle\int_{-t_m}^{\infty}(1-t\,m)\,dP_0(t)}\;,\label{m_r_eq_res_min}\\
    1&=\alpha \frac{\displaystyle\int_{-t_m}^{\infty}\frac{t^2\,m^2}{1-t\, m}\,dP_0(t)}{\displaystyle\int_{-t_m}^{\infty} (1-t\, m)\,dP_0(t)}\;.\label{m_i_eq_res_min}
\end{align}

Note that in the Gaussian model defined by Eqs. (\ref{H}) and (\ref{GRF}) both annealed complexities are universal \cite{fyodorov2004complexity,fyodorov2007density}, depending only on a single real parameter, the ratio $\mu/\sqrt{F''(0)}$. In contrast,  the level of universality  in the class of models considered here is lower, as the annealed complexities explicitly depend on the whole density $p_0(t)$. For a special case of the Gaussian $p_0(t)$ many features of annealed complexities have been analysed in our previous paper \cite{lacroix2022superposition}, but the general case turns out to show richer behaviour.

We now describe in more details the behaviour of these complexities, presenting our main findings as propositions.

\vspace{0.2cm}

\noindent {\bf Proposition 1.} 
{\it 
For any  PDF $p_0(t)$ whose support extends over the whole real line $t\in \mathbb{R}$ and such that $\int |t|\,dP_0(t)<\infty$, the random landscape defined in Eq.(\ref{H}) is always topologically complex for any $\mu>0$, and characterized by non-vanishing annealed complexities: 
\be
\;\Sigma_{\rm tot}(\mu,\alpha)>\Sigma_{\min}(\mu,\alpha)>0\;, \quad \forall \mu\in(0,\infty),
\ee
with both complexities being monotonically decreasing functions of $\mu$.}

\vspace{0.3cm}

 The leading asymptotic behaviour of the complexities for $\mu\to \infty$ can be computed explicitly and reads
\be
\Sigma_{\rm tot}(\mu,\alpha)\approx 2\Sigma_{\min}(\mu,\alpha)\approx 2\alpha\mu\int_{1}^{\infty}(\tau-1)\,p_0(-\mu \tau)\,d\tau\;.
\ee
For a particular choice of a model in this category with $p_0(t)=(2\pi)^{-1/2}\exp(-t^2/2)$ the proposition above has been previously established by us in \cite{lacroix2022superposition}. 

\vspace{0.2cm}

\noindent {\bf Proposition 2.} 
{\it
For any PDF $p_0(t)$ with a finite left edge of its support $-t_{\rm e}$ such that  $p_0(t)=0$ $\forall t<t_{\rm e}$ there exists a finite value $\mu_c(\alpha)$, solution of the equation
\be
\frac{\alpha}{\mu_c(\alpha)}\int \frac{t^2\,dP_0(t)}{\mu_c(\alpha)+t}-1=\alpha\left(\int \frac{\mu_c(\alpha)\,dP_0(t)}{\mu_c(\alpha)+t}-1\right)-1=0\;,\label{mu_c_eq}
\ee
such that the random landscape defined in Eq.(\ref{H})  is topologically trivial in the regime $\mu>\mu_c(\alpha)$, i.e.
\be
\;\;\Sigma_{\rm tot}(\mu,\alpha)=\Sigma_{\min}(\mu,\alpha)=0\;, \quad \forall \mu\geq \mu_c(\alpha)\;.
\ee
Conversely, for any $\mu<\mu_c(\alpha)$ the random landscape is topologically complex, i.e.
\be
\;\;\Sigma_{\rm tot}(\mu,\alpha)>\Sigma_{\min}(\mu,\alpha)>0\;, \quad \forall \mu\in(0,\mu_c(\alpha))\;.
\ee
}

\vspace{0.3cm}

In the vicinity of the topology trivialisation transition value $\mu=\mu_c$ one further finds in the above case that the total annealed complexity and the annealed complexity of minima both vanish quadratically. Explicitly we have:
\begin{align}
\Sigma_{\rm tot}(\mu,\alpha)&= \frac{\alpha(2\mu_c-\alpha I_2(\mu_c))}{2\mu_c\left[\alpha^3 I_2(\mu_c)^2+\mu_c^2+2\alpha\mu_c^2+\alpha^2\mu_c(I_3(\mu_c)-4 I_2(\mu_c))\right]}(\mu-\mu_c)^2+O(\mu-\mu_c)^3\;, \label{tot_edge}\\
\Sigma_{\min}(\mu,\alpha)&= \frac{\alpha (\alpha I_2(\mu_c)-2\mu_c)(\mu_c-t_{\rm e})t_{\rm e}}{2\mu_c^2\left[\mu_c^3+t_{\rm e}\left(\alpha^2 I_2(\mu_c)(\mu_c-t_{\rm e})+(1+\alpha)\mu_c(t_{\rm e}-2\mu_c)\right)\right]}(\mu-\mu_c)^2+O(\mu-\mu_c)^3\;, \label{min_edge}
\end{align}
where we have denoted
\be
I_k(\mu)=\int_{-t_{\rm e}}^{\infty} \frac{t^{1+k}\,dP_0(t)}{(\mu+t)^{k}}\;,
\ee
and we remind that $-t_{\rm e}$ is the left edge of the support for the density $p_0(t)$. See Appendix \ref{ex_model} for an explicit computation of the coefficient in Eq. \eqref{min_edge} for a particular case of our model.\\[1ex]

{\bf Remark}. While quadratic vanishing of the total complexity on approaching the topological trivialization threshold $\mu_c$ is quite standard \cite{fyodorov2004complexity}, the same behaviour for the annealed complexity of minima is somewhat unusual. One would rather expect instead a cubic vanishing of the complexity of minima   \cite{fyodorov2007replica,Fyodorov_Nadal_12,Fyo_2015_rev,BenArousBourgadeMcKenna24}, compatible with the anticipated "third order" nature of the ergodicity breaking transition.
This fact might be an indication of the expression for the annealed complexity of minima in the present model being different from the true quenched complexity of minima, while the latter might be showing cubic vanishing. This feature clearly calls for further investigation of yet outstanding quenched complexity in the present model.\\[0.2ex]

Finally, one can show that both annealed complexities diverge as $\mu\to 0$ with a leading behaviour independent of the PDF $p_0(t)$, that reads
\be
\Sigma_{\rm tot}(\mu,\alpha)\approx \Sigma_{\min}(\mu,\alpha)\approx -\min(1,\alpha)\ln \mu\;,\;\;\mu\to 0\;.
\ee

\subsection{Expression for the ground-state energy}

\noindent {\bf Proposition 3.} 
{\it 
The expected value of the global minimum (the ground-state energy) $e_{0}$ of the random landscape defined in Eqs. (\ref{H}) can be obtained by solving the following optimisation problem for the Parisi-type functional associated with our model: 
\begin{align}
   e_0=\underset{l,w(l')}{\rm sup}\left[\frac{\mu}{2}\int_{0}^{l}\frac{\displaystyle dt}{\displaystyle\left(\mu+\int_{0}^{t} w(\tau)\,d\tau\right)^2}-\frac{1-\alpha}{2}\int_{0}^{l}\frac{\displaystyle dt}{\displaystyle\mu+\int_{0}^{t} w(\tau)\,d\tau}
    -\alpha\ln\mathbb{E}_{\phi}\left[f(0,0)\right]\right]\;,\label{gen_opt}
\end{align}
where supremum is taken over both the real parameter $l\ge 0$ and all the non-negative functions $w(\tau)$ non-decreasing in the interval $\tau\in[0,l]$, while the expectation $\mathbb{E}_{\phi}\left[\cdots\right]$ is taken with respect to the distribution of the random function $\phi(z)$. The function $f(t,h)$ entering the last term of the optimisation functional satisfies the Parisi partial differential equation
\be\label{Parisi_Diff_EqA}
\partial_t f=-\frac{1}{2}\left[\partial_h^2 f+w(t)\left(\partial_h f\right)^2\right]\;,
\ee
with the boundary condition which explicitly depends on a realization of the random function $\phi(z)$ as
\be
f(t\geq l,h)=-\epsilon_{\min}\left(\mu+\int_{0}^{l} \,w(\tau)\,d\tau,h\right)\;, \label{bc}
\ee
where the function of two variables $\epsilon_{\min}(\nu,h)$ in Eq. (\ref{bc}) is given by 
\be
\epsilon_{\min}(\nu,h)=\min_z H_{\nu,h}(z)\;,\;\;H_{\nu,h}(z)=\frac{\nu}{2}z^2-h z+\phi(z)\;.\label{H_e_min}
\ee
 
}

\vspace{0.3cm}

One may think of  $\epsilon_{\min}(\nu,h)$  as the ground-state energy of the one-dimensional disordered Hamiltonian $H_{\nu,h}(z)$. Let us denote the position of the corresponding minimum
\be
z_{\min}(\nu,h)=\underset{z}{\rm argmin}\, H_{\nu,h}(z)\;,
\ee
and similarly $z_{\min}(\nu)\equiv z_{\min}(\nu,0)$. It will be convenient further to introduce the functions ${\cal C}_k(\nu)$ according to
\be
\xi(\nu,h)=\lim_{\beta\to \infty}\frac{1}{\beta}\ln\frac{\int e^{-\beta H_{\nu,h}(z)}\,dz}{\int e^{-\beta H_{\nu,0}(z)}\,dz}=\epsilon_{\min}(\nu,0)-\epsilon_{\min}(\nu,h)=\int_{0}^h z_{\min}(\nu,h_0) \,dh_0\,=\sum_{k=1}^{\infty}\frac{h^k}{k!}{\cal C}_k(\nu)\;.\label{C_k_cum}
\ee
Such functions will play a significant role in the following and can be interpreted as rescaled "thermal" cumulants of the position $z$, in the limit of zero temperature. Note however that these cumulants are still random variables as they depend on the specific realisation of the disordered potential $\phi(z)$. 

Depending on the range of parameters $\alpha$, $\mu$ and statistics of $\phi(z)$ the optimiser $ w(\tau)$ and  the length $l$ of its interval of support for the ground state value Parisi-type functional may be of different types,
  the simplest being the replica-symmetric (RS) and then the one-step replica symmetry broken (1RSB) solution. The transitions between RS, 1RSB and eventually full replica symmetry broken (FRSB) types may also be of different nature, and we proceed with briefly describing this rich picture below.

{\bf Remark.} In the standard theory of RSB, an optimal (rescaled) Parisi function $w(\tau)$ describing a phase with  $k$-step RSB pattern displays $k+1$ plateau regions, see e.g.\cite{auffinger2020sk}. In the case studied here, we find it more convenient to define the Parisi function $w(\tau)$  in a slightly different manner such that its $k$ plateaus describe a phase with $k$-step RSB pattern.

\subsubsection{Replica-symmetric solution}

In the replica-symmetric solution, the expression for the mean ground-state energy $e_0$ in Eq. \eqref{gen_opt} is obtained by choosing the parameter $l$ ( i.e. the length of the interval supporting $w(\tau)$)  to be zero (or, equivalently, $w(\tau)\equiv 0$) and eventually reads
\be
 e_0=\alpha\,\Ep{\epsilon_{\min}(\mu)}:=e_{\rm RS}\;.
\ee
where we denoted $\epsilon_{\min}(\nu)=\epsilon_{\min}(\nu,0)$.

Introducing the self-overlap $r_d$ and the overlaps $r_{ab}$, defined as  
\be
r_d=\frac{{\bf x}_a^2}{N}\;,\;\;r_{ab}=\frac{{\bf x}_a\cdot{\bf x}_b}{N}\;,\;\;a\neq b\;,
\ee
the latter  values in the ground state in the replica-symmetric phase are independent of $(a,b)$ and will be denoted as $r$. Explicitly, they satisfy in the zero-temperature limit the following relations:

\be
r_0=\lim_{\beta\to \infty}r=\lim_{\beta\to \infty}r_d=\alpha\,\Ep{z_{\min}^2(\mu)}\;,\;\;v=\lim_{\beta\to \infty}\beta(r_d-r)=\frac{1}{\mu}\;.
\ee

\subsubsection{Continuous transition to the replica symmetry broken solution}

The validity of the RS solution  $l=0$ for the ground state optimisation problem is ensured by the corresponding "replicon" eigenvalue being negative:
\be\label{replicon_eigv}
\lambda_{\rm RS}=\alpha\mathbb{E}_{\phi}\left[\frac{\mu^2}{\left(\mu+\phi''[z_{\min}(\mu)]\right)^{2}}-1\right]-1\leq 0\;.
\ee
When such a negativity is violated, a continuous transition to the replica symmetry broken (RSB) solution may occur along the corresponding de Almeida-Thouless (AT) line\cite{AT78}. The  position of the AT line is determined by the AT criterion $\lambda_{\rm RS}=0$, and in the range of system parameters where $\lambda_{\rm RS}>0$ the solution is always RSB, i.e. requires $l>0$.
In the Appendix \ref{match_crit} we further show that the AT criterion for the RSB transition matches exactly the criterion for the topology trivialisation transition given in Eq. \eqref{mu_c_eq}. Although both criteria reflect the occurrence of ergodicity breaking it is apriori not clear they should match as the Eq. \eqref{mu_c_eq} was based on annealed complexity, hence only may provide a bound. The matching between these criteria for the present model is therefore worth noting. It makes us to conjecture that, when the corresponding RSB transition is continuous, the yet unknown quenched complexity should vanish at the same  values of parameters where the annealed complexity vanishes, sharing this property with the Gaussian toy model defined by Eqs. (\ref{H}-\ref{GRF}) where such matching was previously observed in  \cite{fyodorov2007replica}, 
and with the spherical model  \cite{belius2022triviality,Huang_Sellke23}. On the other hand, as we already discussed above in the present model the value of the quenched and annealed complexity function may well be different everywhere in the RSB phase.

The properties of the RSB phase can be investigated further. At the transition, the solution $w(\tau)$ starts to differ from its RS expression $w(\tau)=0$ value. Such a difference is described by the value of the so-called  "breaking point" \cite{parisi_urbani_zamponi2020book} which can be found via a procedure described above Eq. \eqref{breaking_point2} of the main text, and is given for our model as
\be
w_{\rm AT}=\frac{\alpha\mu^3\Ep{{\cal C}_3(\mu)^2}}{2\left(\alpha\mu^3\Ep{{\cal C}_2(\mu)^3}-(\alpha+2)\right)}\;,
\ee
where the functions ${\cal C}_k$'s are defined in Eq. \eqref{C_k_cum}, and in particular ${\cal C}_2(\mu)=\left(\mu+\phi''[z_{\min}(\mu)]\right)^{-1}$. In case such a breaking point is negative, the transition from RS phase to the RSB phase is expected to be discontinuous rather than continuous.

Following the approach of \cite{parisi_urbani_zamponi2020book}, one may further infer the nature of the RSB phase  in the vicinity of the transition. It is controlled by the sign of the following quantity
\be
w_{\rm AT}'=\frac{\alpha\mu^4\left(\mathbb{E}\left[{\cal C}_4(\mu)^2\right]-12 w_{\rm AT}\,\mathbb{E}\left[{\cal C}_3(\mu)^2{\cal C}_2(\mu)\right]+6w_{\rm AT}^2\,\mathbb{E}\left[{\cal C}_2(\mu)^4\right]\right)-6(\alpha+3)w_{\rm AT}^2}{2\left(\alpha\mu^3\,\mathbb{E}\left[{\cal C}_2(\mu)^3\right]-(\alpha+2)\right)}\;,
\ee
reflecting the condition of $w(\tau)$ being non-decreasing. 
Namely, the continuous transition from RS to RSB is actually towards a phase with one-step replica symmetry breaking (1RSB) if $w_{\rm AT}'<0$. In the latter case $l>0$ and the minimizer $w(\tau)$ has a constant value   
$w(t)=m$ for $0\le t\le l$. In the opposite case $w_{\rm AT}'>0$ the transition is towards the full replica symmetry breaking solution (FRSB) characterized by a nontrivial function $w(\tau)$ increasing in the interval $[0,l]$, see  \cite{parisi_urbani_zamponi2020book} for a detailed discussion. In fact  more exotic RSB types can not be excluded at some special values of parameters, but won't be further discussed,   see \cite{AufZeng2019}
for such examples in the framework of spherical model.

\subsubsection{Random first order transition}

In addition to the continuous transition from the RS phase to a RSB phase, which may be FRSB or 1RSB, a discontinuous transition may occur between the RS phase and a 1RSB phase. Such transition is sometimes called in the literature the "random first order transition" (RFOT) \cite{biroli2012random}. The location of such phase transition cannot however be obtained by analysing the replicon eigenvalues Eq. \eqref{replicon_eigv}. In contrast, one needs to  compute explicitly the sign of the ground-state energy difference $\Delta e_{\rm 1RSB}$ between the values at RS and at 1RSB phases. The latter is computed by exploiting the 1RSB expression for the minimizer of the Parisi-type functional Eq. \eqref{gen_opt} which amounts to replacing 
$w(t)=m$ for $t<l$ in that equation. The procedure yields the required energy difference in the form
\begin{eqnarray}
     \Delta e_{\rm 1RSB}=\underset{l,m\geq 0}{\rm sup}\left\{\frac{1}{2}\left(\frac{l}{\mu+ml}-\frac{1-\alpha}{m}\ln\left(1+\frac{m l}{\mu}\right)\right)\right. \nonumber \\ \left. -\frac{\alpha}{m}\mathbb{E}\left[\ln\left(\int e^{-\frac{h^2}{2l}-m\epsilon_{\min}(\mu+m l,h)}\frac{dh}{\sqrt{2\pi l}}\right)+m\epsilon_{\min}(\mu)\right]\right\}\;.\label{d1RSB}
\end{eqnarray}

Let us denote $l_*$ and $m_*$ the optimal value of $l$ and $m$ in this optimisation problem. The expression above for $\Delta e_{\rm 1RSB}$ vanishes as $l_*\to 0$. Expanding the ground-state energy in  powers of $l_*$ one may then check that the first order in such an expansion vanishes, while the corresponding coefficient in front of the second order term reproduces the expression of the replicon eigenvalue (see Eq. \eqref{replicon_eigv}).  This procedure provides  an alternative method to derive the criterion for the continuous replica symmetry breaking transition. 

 Similarly, using the identity $\Ep{\epsilon_{\min}(\mu,h)}=\Ep{\epsilon_{\min}(\mu)}-\frac{h^2}{2\mu}$, one can check that $\Delta e_{\rm 1RSB}$ vanishes also as $m_*\to 0$.  Expanding now Eq. \eqref{d1RSB} up to the first order in $m_*$ one arrives at $\Delta e_{\rm 1RSB}=A(l)m_*+O(m_*^2)$, where 
\begin{align}
    A(l)=&\frac{\alpha}{2}\,\Ep{\left(\int e^{-\frac{h^2}{2l}}\epsilon_{\min}(\mu,h)\frac{dh}{\sqrt{2\pi l}}\right)^2-\int e^{-\frac{h^2}{2l}}\left[\epsilon_{\min}^2(\mu,h)-l\,z_{\min}^2(\mu,h)\right]\frac{dh}{\sqrt{2\pi l}}}\\
    &-\frac{l^2(1+\alpha)}{4\mu^2}\;.\nn
\end{align}
The location of the RFOT is obtained by ensuring that the ground-state energy difference vanishes at a non-trivial value of $l_*$ by requiring simulatneously $A(l_*)=0$ and $A'(l_*)=0$ for $l_*\ne 0$. Note that the trivial solution $l_*=0$ always exists and recovers again the de Almeida-Thouless line.

\subsubsection{Gardner transition}

So far, we have only considered transitions between the RS phase and a RSB phase, be it 1RSB or FRSB. A Gardner transition \cite{gardner1985spin} between these two qualitatively different RSB phases may additionally arise in phase-space. In full similarity with the criterion of stability for the RS phase in terms of the negativity of the replicon eigenvalue, one can 
formulate the criterion of 1RSB phase being stable. To achieve this we first introduce 
the probability density $P(l,h)$ via
\be
P(l,h)=\frac{\displaystyle e^{-\frac{h^2}{2l}-m \epsilon_{\min}(\mu+m l,h)}}{\displaystyle \int e^{-\frac{h_0^2}{2l}-m \epsilon_{\min}(\mu+m l,h_0)}dh_0}\,\;.
\ee
where the parameters $m$ and $l$ characterizing 1RSB solution take their actual values obtained by solving the optimisation problem in Eq. \eqref{d1RSB}.
Further denote $\epsilon_{\min}\equiv \epsilon_{\min}(\mu+m l,h)$ and $P(l,h)\,dh: =d{\cal P}_l(h)$. Then one finds that  1RSB phase
is only stable if the largest of the two following eigenvalues is negative:
\begin{align}
\lambda_1=\alpha\mu^2 \Ep{\left(\int \,\partial_{h}^2\epsilon_{\min}\, d{\cal P}_l(h)
-  m\left[\int \,(\partial_{h}\epsilon_{\min})^2\,d{\cal P}_l(h) - \left(\int 
\,\partial_{h}\epsilon_{\min}\,d{\cal P}_l(h)\right)^2\right]\right)^2}
 -1+\alpha 
 \end{align}
 and 
 \begin{eqnarray}
\lambda_2=\alpha\mu^2\Ep{\int \left(\partial_{h}^2\epsilon_{\min}\right)^2\,d{\cal P}_l(h) -
\frac{2\mu^3}{(\mu+m l)^3}+\frac{(1-\alpha)\mu^2}{(\mu+m l)^2}}\;.
\end{eqnarray}

\section{Derivation of the annealed complexities} \label{sec_complexity}

In this section, we provide detailed derivations for the main results on the numbers of critical points and minima and their associated complexity, presented in the previous section. We also provide additional analysis of these results, in particular on some properties of the optimal Hessian spectrum.

\subsection{Total number of critical points}

We start with considering the expectation of the total number of critical points given via the Kac-Rice formula in the form 
\be
\mathbb{E}\left[{\cal N}_{\rm tot}(\mu)\right]=\int d{\bf x}\,\mathbb{E}\left[\prod_{i=1}^N\delta\left(\mu x_i+\sum_{l=1}^M k_{li}\phi_l'({\bf k}_l\cdot{\bf x})\right)\left|\det_{1\leq i,j\leq N}\left(\mu\delta_{ij}+\sum_{l=1}^M k_{li}k_{lj}\phi_l''({\bf k}_l\cdot{\bf x})\right)\right|\right]\;.
\ee
Let us introduce convenient notations for further use, denoting the random variables
\be
    G_l=\phi_l'({\bf k}_l\cdot{\bf x})\;,\;\;
    T_l=\phi_l''({\bf k}_l\cdot{\bf x})\;,\;\;l=1,\cdots,M\;.
\ee
The random variables $T_l$'s are  i.i.d., and similarly the variables $G_l$'s, but  for a fixed index $l$ the random variables $T_l$ and $G_l$ may be correlated. We denote $p_0(t)$ the  probability density function (PDF) of each of the random variables $T_l$. It is convenient further to introduce two matrices: the matrix $K$ of size $N\times M$, whose columns correspond to the $M$ wave vectors ${\bf k}_l$, and the diagonal matrix $T$ with entries satisfying $T_{kl}=T_l \delta_{kl}$. These definitions imply 
\be
\mu\delta_{ij}+\sum_{l=1}^M k_{li}k_{lj}\phi_l''({\bf k}_l\cdot{\bf x})=\left(\mu\mathbb{I}+K T K^T\right)_{ij}\;,\;\;i,j=1,\cdots,N\;.
\ee
One further notices that the stationarity of the i.i.d. random functions $\phi_l(z)$'s, i.e. the invariance of their statistics by translations $z\to z+a$ implies that the statistics of both random variables $G_l$ and $T_l$ is independent of the position ${\bf x}$ and the wavevector ${\bf k}_l$. This fact allows to exchange safely the integration over ${\bf x}$ and the expectation, giving
\be
\mathbb{E}\left[{\cal N}_{\rm tot}(\mu)\right]=\mathbb{E}\left[\int d{\bf x}\,\prod_{i=1}^N\delta\left(\mu x_i+\sum_{l=1}^M k_{li}G_l\right)\left|\det\left(\mu\mathbb{I}+K T K^T\right)\right|\right]\;.
\ee
Now the spatial integral over ${\bf x}$ can be immediately evaluated due to the $\delta$-function factors, yielding a trivial factor $\mu^{-N}$. This brings the mean number of critical points  for any value of $N$ and $M$ to the form
\begin{align}
    \mathbb{E}\left[{\cal N}_{\rm tot}(\mu)\right]&=\frac{1}{\mu^N}\mathbb{E}\left[\left|\det\left(\mu\mathbb{I}+K T K^T\right)\right|\right]\;,\label{n_tot}
\end{align}
where the expectation is taken with respect to the random wavevectors ${\bf k}_l$'s and the i.i.d. random variables $T_l$'s. 

Our first aim is to evaluate the total annealed complexity
\be
\Sigma_{\rm tot}(\mu,\alpha)=\lim_{N,M\to \infty}\frac{1}{N}\ln \mathbb{E}\left[{\cal N}_{\rm tot}(\mu)\right]\;.
\ee
  in the high-dimensional limit $N,M\to \infty$ keeping $\alpha=M/N=O(1)$ finite. When extracting the required asymptotic behaviour of the determinant in \eqref{n_tot} we will assume that the columns ${\bf k}_l$'s for $l=1,\cdots,M$ of the matrix $K$ are either independent Gaussian random vectors or independent uniformly distributed vectors on the $N$-sphere ${\bf k}\in \sqrt{N}{\cal S}_{N-1}$. We similarly suppose that the elements $T_l$'s of the diagonal matrix $T$ are independent and identically distributed with zero mean, and with the  density $p_0(t)$ satisfying $\int |t|\,p_0(t)\,dt<\infty$. Under these conditions we expect the 
following identity of limits
\be
\lim_{N,M\to \infty}\frac{1}{N}\ln\mathbb{E}\left[\det\left|\mu\mathbb{I}+K T K^{T}\right|\right]=\lim_{N,M\to \infty}\frac{1}{N}\mathbb{E}\left[\ln\det\left|\mu\mathbb{I}+K T K^{T}\right|\right]=\int \rho(\lambda)\ln|\mu+\lambda|\,d\lambda\,\label{strong_self_av}
\ee
to hold. Here $\rho(\lambda)=\lim_{N,M\to \infty}N^{-1}{\rm Tr}\left[\delta(K T K^{T}-\lambda\mathbb{I})\right]$ is the limiting spectral density of the random matrix $K T K^{T}$. The validity of \eqref{strong_self_av} is natural to call the {\bf strong self-averaging property}.  Such  property has been proved to hold for several large classes of random determinants considered in \cite{BenArousBourgadeMcKenna23}, and cover, in particular, the special case of matrices  $K T K^{T}$ with positive definite  $T>0$. Although the case of the matrices $K T K^{T}$  with $T_l$'s of any sign, most relevant to the present context, seems to go beyond the remit of theorems proved in   \cite{BenArousBourgadeMcKenna23}, it is very natural to conjecture their validity extends to such a case as well. \\[1ex]
To obtain our final result for the annealed total complexity, we introduce a change of variables from the set of random variables $T_l$'s to the limiting spectral density $p(t)$ of the matrix $T$, which incurs a Jacobian term in the process. We find it most economic to follow the functional integral form of the Large Deviation approach as exposed  e.g. in\cite{DeanMajumadar2008extreme}. After standard manipulations, details of which can be found in our previous paper \cite{lacroix2022superposition} (see eqs. (38)-(48) there) the annealed complexity is found to be given by
\be
\Sigma_{\rm tot}(\mu,\alpha)=\max_{p(t):\int p(t)\,dt=1}\left[-\alpha\int p(t)\,\ln\left(\frac{p(t)}{p_0(t)}\right)\,dt+\int \rho(\lambda)\ln\left|\frac{\mu+\lambda}{\mu}\right|\,d\lambda\right]\;,\label{sig_tot_opt}
\ee
where the first term having the form of the Kullback-Leibler divergence stems from the Jacobian of the functional transformation. Further progress is possible after employing the original results from the seminal work of Marchenko and Pastur \cite{pastur1967distribution}, allowing to express explicitly the dependence of the limiting spectral density $\rho(\lambda)$ of the random matrix $K T K^{T}$ in terms of the spectral density $p(t)$ of the matrix $T$. These results are best stated by introducing the Stieltjes transform of the limiting density
\be
m(z;\alpha)=\int \frac{\rho(\lambda')}{z-\lambda'}d\lambda'\,\;,\;\;z\in \mathbb{C}\;.\label{m_def}
\ee
For $\lambda\in \mathbb{R}$, the Stieltjes transform reads $m(\lambda;\alpha)=m_r(\lambda;\alpha)+i\,m_i(\lambda;\alpha)$, and the limiting density is expressed exactly as $\rho(\lambda)=m_i(\lambda;\alpha)/\pi$. Marchenko and Pastur proved that the Stieltjes transform satisfies the following self-consistent equation \cite{pastur1967distribution}
\be
\frac{1}{m(z;\alpha)}=z-\alpha\,\int \frac{t\,p(t)}{1-t\,m(z;\alpha)}\,dt\;,\;\;z\in \mathbb{C}\;,\;\;\alpha\in\mathbb{R}_+\;.\label{MP_m_eq}
\ee
The variation of the second term in the optimisation functional Eq. \eqref{sig_tot_opt} (which we for brevity will call the "action" below) can be performed explicitly by using the following identity
\be \label{identderiv}
\lim_{\epsilon\to 0}\Re\left[\int \partial_{\nu}\ln\frac{\nu+i \epsilon+\lambda}{\nu}\, \rho(\lambda) d\lambda\right]\,=\dashint \frac{\rho(\lambda)}{\nu+\lambda} d\lambda-\frac{1}{\nu}=-m_r(-\nu;\alpha)-\frac{1}{\nu}\;,
\ee
as well as another identity which may be simply derived from Eq. \eqref{MP_m_eq} upon variation:
\be
\frac{\delta m (z;\alpha)}{\delta p(t)}=\alpha\,\partial_z \ln(1-t\,m(z;\alpha))\;.
\ee
Using them one may now compute explicitly the functional derivative of the functional to be optimised in  Eq. \eqref{sig_tot_opt}  over the normalised limiting density $p(t)$. 
The stationarity condition yields the following relation for the optimal density $p_*(t)$:
\begin{align}
 \ln \left(\frac{p_*(t)}{Z_p\,p_0(t)}\right)&=\frac{1}{\alpha}\int d\lambda\left.\frac{\delta \rho(\lambda)}{\delta p(t)}\right|_{p(t)=p_*(t)}\ln\left|\frac{\mu+\lambda}{\mu}\right|=\frac{1}{\alpha}\int_{\mu}^{\infty}d\nu\frac{\delta }{\delta p(t)}\left(m_r(-\nu;\alpha)+\frac{1}{\nu}\right)_{p(t)=p_*(t)}\\
 &=\ln |1-t\,m(-\mu;\alpha)|\;,  \nn 
\end{align}
where $Z_p$ is a normalization factor ensuring that $\int p_*(t)\,dt=1$. The optimal density can thus be obtained explicitly for a fixed value of the Stieltjes transform $m(-\mu;\alpha)$ as 
\be
p_*(t)=\frac{\displaystyle p_0(t)|1-t\,m(-\mu;\alpha)|}{\displaystyle\int \,p_0(t')|1-t'\,m(-\mu;\alpha)|\,dt'}\;.
\ee
Finally, given this limiting density, and denoting from now on $p_0(t)dt:=dP_0(t)$,  the value of the Stieltjes transform $m(-\mu;\alpha)$ needs to be obtained self-consistently by inserting this optimal density in Eq. \eqref{MP_m_eq} for $z=-\mu$, yielding
\be
\frac{1}{m(-\mu;\alpha)}=-\mu-\alpha\frac{\displaystyle\int\, t\frac{|1-t\,m(-\mu;\alpha)|}{1-t\,m(-\mu;\alpha)}\,dP_0(t)}{\displaystyle \int \,|1-t'\,m(-\mu;\alpha)|\,dP(t')}\;.\label{m_eq}
\ee
Taking the real and imaginary part of the above equation, one can obtain more explicit equations for the real part $m_r\equiv m_r(-\mu;\alpha)$ of the Stieltjes transform which will be instrumental in the computation of the total annealed complexity and the imaginary part $m_i\equiv m_i(-\mu;\alpha)=\pi\rho(-\mu)$ providing the limiting eigenvalue density of the matrix  $K T K^T$ at the spectral point $-\mu$. These equations read
\begin{align}
\frac{m_r}{m_r^2+m_i^2}&=\mu-\alpha\frac{\displaystyle\int 
\frac{t\,\,(1-t\,m_r)}{\displaystyle\sqrt{(1-t\,m_r)^2+t^2 m_i^2}}\,dP_0(t)}{\displaystyle \int \,\sqrt{(1-t'\,m_r)^2+t'^2 m_i^2}\,dP(t')}\;,\\
\frac{m_i}{m_r^2+m_i^2}&=\alpha m_i\frac{\displaystyle\int \frac{t^2}{\displaystyle\sqrt{(1-t\,m_r)^2+t^2 m_i^2}}\,dP_0(t)}{\displaystyle \int \sqrt{(1-t'\,m_r)^2+t'^2 m_i^2}\,dP(t')}\;.\label{eq_mi}    
\end{align}
The only dependence of the spectral density $\rho(\lambda)$ on the parameter $\mu$ is implicit and comes from its dependence on the optimiser $p_*(t)$. In order to analyse the total complexity in Eq. \eqref{sig_tot_opt}, we introduce a simplified expression for the latter, obtained by first computing its derivative with respect to $\mu$, which reads
\begin{align}
 \partial_{\mu}\Sigma_{\rm tot}(\mu,\alpha)&=-\alpha\int dt\,\partial_{\mu}\left(p_*(t)\,\ln\frac{p_*(t)}{p_0(t)}\right)+\int d\lambda\,\partial_{\mu}\left(\rho(\lambda)\ln\left|\frac{\mu+\lambda}{\mu}\right|\right)\\
 &=\int dt\,\partial_{\mu}p_*(t)\left[-\alpha\ln \frac{p_*(t)}{p_0(t)}+\int d\lambda\left.\frac{\delta \rho(\lambda)}{\delta p(t)}\right|_{p(t)=p_*(t)}\ln\left|\frac{\mu+\lambda}{\mu}\right|\right]+\dashint \frac{d\lambda\,\rho(\lambda)}{\mu+\lambda}-\frac{1}{\mu}\;.\nn
\end{align}
The first integral in the second line above vanishes due to stationarity condition, yielding with the help of Eq. \eqref{identderiv} the following simple expression
\be
\Sigma_{\rm tot}(\mu,\alpha)=\int_{\mu}^{\infty}\left(m_r(-\nu;\alpha)+\frac{1}{\nu}\right)\,d\nu\;.\label{sig_tot-form}
\ee
The integrand in the above can be re-expressed using Eq. \eqref{m_eq} as
\be
m_r(-\nu;\alpha)+\frac{1}{\nu}=\frac{\alpha}{\nu}\frac{\displaystyle\int |1-t\,m|\left(1-\frac{1-t\,m_r}{\sqrt{(1-t\,m_r)^2+t^2 m_i^2}}\right)\,dP_0(t)}{\displaystyle \int \,|1-t'\,m|\,dP(t')}\geq 0\;.\label{integrand}
\ee
 Clearly $\frac{1-t\,m_r}{\sqrt{(1-t\,m_r)^2+t^2 m_i^2}}<1$  in the case where $m_i=\pi\rho(-\nu)>0$. In addition, even if $m_i=\pi\rho(-\nu)=0$, to ensure that the expression in Eq. \eqref{integrand} vanishes one needs $1-t\, m_r>0$ for any $t$ within the support of the PDF $p_0(t)$ . The latter clearly will be violated for $t<1/m_r$, and as $m_r<0$ can not hold for any PDF $p_0(t)$ with unbounded support on the left. As the integrand in the expression of the total annealed complexity in Eq. \eqref{sig_tot-form} is positive for any finite value of $\nu>0$, we conclude that the complexity is strictly positive and decreasing for any finite value of $\mu$. Conversely, for a PDF $p_0(t)$ with a finite left edge $t_{\rm e}$, the integrand in Eq. \eqref{integrand} vanishes only if both $m_i=\pi \rho(-\nu)=0$ and $m_r=-\nu^{-1}>-1/t_{\rm e}$. One can show that more stringent criterion is the former and it is operative for any $\nu>\mu_c(\alpha)$, where the value $\mu_c(\alpha)$ is obtained by ensuring that $-\mu_c(\alpha)$ lies precisely at the left edge of the limiting distribution $\rho(\lambda)$. The associated equation is derived from Eq. \eqref{eq_mi} by requiring that the non-trivial solution coincides with the trivial one $m_i=0$. It reads
\be
\mu_c(\alpha)=\alpha\int \frac{t^2}{\mu_c(\alpha)+t}\,dP_0(t)=\alpha\left[\int 
\frac{\mu_c(\alpha)^2\,}{\mu_c(\alpha)+t}\,dP_0(t)-\int \left(\mu_c(\alpha)-t\right)\,dP_0(t)\right]
\ee
\be
=\alpha\mu_c(\alpha)\left[\int \frac{\mu_c(\alpha)}{\mu_c(\alpha)+t}\,dP_0(t)-1\right]\;.
\ee
As the integrand in Eq. \eqref{sig_tot-form} vanishes for any $\nu>\mu_c(\alpha)$ and is strictly positive otherwise, one may conclude that in the regime $0<\mu<\mu_c(\alpha)$ the total annealed complexity is strictly positive and decreasing while for $\mu>\mu_c(\alpha)$ it is constant and equal to zero.

Let us now analyse the asymptotic behaviour of the total annealed complexity in the limit of small curvature of the confining parabolic potential $\mu\to 0$. The leading behaviour in that case turns out to be independent of the density $p_0(t)$ but depends only on the value of $\alpha$. In particular, for $\alpha>1$ (i.e. $M>N$), the matrix $K T K^T$ is full rank and one expects that $m_r\to 0$ as $\nu \to 0$ while $m_i$ reaches a finite value. Supposing that the limiting density is symmetric: $p_0(t)=p_0(-t)$ one may check in this case that at the leading order the equations determining $m_r,m_i$ read
\be
\frac{m_r}{m_i^2}=-\nu\;,\;\;\frac{1}{\alpha}=\frac{\displaystyle \int \frac{t^2 m_i^2}{\sqrt{1+t^2 m_i^2}}\,dP_0(t)}{\displaystyle \int \sqrt{1+t^2 m_i^2}\,dP_0(t)}\;,\;\;\nu\to 0\;,\;\;\alpha>1\;.
\ee
The second equation provides the value of $m_i=\pi\rho(0)$, which is a decreasing function of $\alpha$, diverging as $\alpha\to 1$ and vanishing as $\alpha\to \infty$ (with $\rho(0)\approx \pi^{-1}\left(\alpha\int dt\,t^2\,p_0(t)\right)^{-1/2}$ if $p_0(t)$ has a finite second moment). The asymptotic behaviour of the complexity reads in that regime 
\be
\Sigma_{\rm tot}(\mu,\alpha)\approx -\ln \mu\;,\;\;\mu\to 0\;,\;\;\alpha>1\;
\ee
In the regime $0<\alpha<1$, the matrix $K T K^{T}$ is of rank $M<N$, and its limiting spectral density thus displays a delta peak $(1-\alpha)\delta(\lambda)$ in addition to its continuous spectrum. The leading behaviour of the Stieltjes transform is determined by this delta peak and reads
\be
m(-\nu;\alpha)\approx -\frac{1-\alpha}{\nu}+m_0(\alpha)\;,\;\;\nu\to 0\;.
\ee
One may check from Eq. \eqref{m_eq} that the only consistent solution for the density $\rho(0)=m_i/\pi$ is zero in the regime $0<\alpha<1$. From the behaviour of the Stieltjes transform, one may simply check that the complexity behaves as
\be
\Sigma_{\rm tot}(\mu,\alpha)\approx -\alpha\ln \mu\;,\;\;\mu\to 0\;,\;\;0<\alpha<1\;.
\ee

Let us now consider the behaviour of the complexity on approaching those values of $\mu$ where the complexity vanishes. One clearly needs to distinguish between the case of PDFs $p_0(t)$ with unbounded support, for which the complexity only vanishes as $\mu\to \infty$ and the case of bounded support for which the complexity vanishes at a finite value $\mu=\mu_c(\alpha)$.

In the former case, one may naturally expect that the density $\rho(-\nu)$ decays fast to zero as $\nu\to \infty$ and we suppose in particular that $m_i\ll m_r$ in that regime. Using this approximation, one may compute
\begin{align}
m_r+\frac{1}{\nu}=dm_r&\approx \frac{\alpha}{\nu^2}\int dt\,t\, p_0(t)\,{\rm sign}\left(1+\frac{t}{\nu}\right) =-\frac{2\alpha}{\nu^2}\int_{-\infty}^{-\nu} dt\,t\, p_0(t)=\frac{2\alpha}{\nu^2}\int_{\nu}^{\infty} dt\,t\, p_0(-t)\\
&=2\alpha \int_{1}^{\infty} dx\,x\, p_0(-\mu x)\;.\nn
\end{align}
Performing an integration by parts, one may re-express the complexity as
\begin{align}
  \Sigma_{\rm tot}(\mu,\alpha)&\approx 2\alpha\int_{\mu}^{\infty}\frac{d\nu}{\nu^2}\int_{\nu}^{\infty} dt\,t\, p_0(-t)=2\alpha\left(\left[-\frac{1}{\nu}\int_{\nu}^{\infty} dt\,t\, p_0(-t)\right]_{\mu}^{\infty}-\int_{\mu}^{\infty}d\nu\,
  p_0(-\nu)\right)\nn\\
  &=2\alpha\int_{\mu}^{\infty} dt\,\left(\frac{t}{\mu}-1\right)\, p_0(-t) = 2\alpha\mu\int_{1}^{\infty} dx\,\left(x-1\right)\, p_0(-\mu x)\;.\label{sig_tot_mu_large}
\end{align}
One may consider a particular example supposing that the bare density $p_0(-t)\approx B t^{\eta}e^{-\omega t^{\chi}}$  with $\chi>0$, so that it vanishes faster than any power law as $t\to \infty$. The associated complexity then behaves as  $\Sigma_{\rm tot}(\mu,\alpha)\approx 2\alpha B/(\chi\omega )^2\mu^{1+\eta-2\chi}e^{-\omega \mu^{\chi}}$ as $\mu\to\infty$. If instead the bare density vanishes with a power law $p_0(-t)\approx B t^{-\eta}$ with $\eta>2$, the complexity vanishes as $\Sigma_{\rm tot}(\mu,\alpha)\approx 2\alpha B/[(\eta-1)(\eta-2)]\mu^{1-\eta}$ as $\mu\to\infty$.

In the second case of the limiting PDF $p_0(t)$ with a bounded support, the behaviour of the complexity in the vicinity of the topology trivialisation transition can be obtained by computing its successive derivatives for $\mu=\mu_c(\alpha)$. In particular, one can simply show that
\be
\partial_{\mu}\Sigma_{\rm tot}(\mu_c(\alpha),\alpha)=-\left(m_r(-\mu_c(\alpha);\alpha)+\frac{1}{\mu_c(\alpha)}\right)=0\;.
\ee
The next derivative yields instead
\be
\partial_{\mu}^2\Sigma_{\rm tot}(\mu_c(\alpha),\alpha)=-\partial_{\mu}m_r(-\mu_c(\alpha);\alpha)+\frac{1}{\mu_c(\alpha)^2}\;.
\ee
Using Eqs. \eqref{sig_tot-form} 
and \eqref{integrand} , the expression above reads
\begin{align}
\partial_{\mu}^2\Sigma_{\rm tot}(\mu_c(\alpha),\alpha)=&-\frac{\mu_c(\alpha)}{2}\partial_{\mu} m_i^2(-\mu_c(\alpha);\alpha)\\
=&\frac{\alpha(2\mu_c-\alpha I_2(\mu_c))}{\mu_c\left[\alpha^3 I_2(\mu_c)^2+\mu_c^2+2\alpha\mu_c^2+\alpha^2\mu_c(I_3(\mu_c)-4 I_2(\mu_c))\right]}\;,\nn
\end{align}
where we have defined
\be
I_k(\mu)=\int \frac{t^{1+k}}{(\mu+t)^{k}}\,dP_0(t)\;,
\ee
with $\alpha I_1(\mu_c)=\mu_c$ in particular. This computation leads to the quadratic threshold behaviour for the total
annealed complexity, Eq. \eqref{tot_edge}.
\vspace{0.3cm}

{\it Some properties of the spectral density for the Hessian at stationary points of the landscape}

Let us now analyse some features of the conditioned Hessian eigenvalue density $\rho(\lambda)$ generated by the optimal density $p_*(t)$. Recall, that this corresponds to studying spectral density of Hessian eigenvalues 
averaged only over all stationary points of the random potential Eq. \eqref{H}, rather than at a generic point in space.  We denote $g_r\equiv m_r(\lambda;\alpha)$ and $g_i\equiv m_i(\lambda;\alpha)$ the real and imaginary part of the Stieltjes transform at the spectral point $\lambda\in \mathbb{R}$. For a given value of $\mu$ that fixes the density $p_*(t)$ the parameters $g_r$ and $g_i$ satisfy
\begin{align}
\frac{g_r}{g_r^2+g_i^2}&=\lambda-\alpha\int \frac{t\,(1-t\,g_r)}{(1-t\,g_r)^2+t^2 g_i^2}\,dP_*(t)\;,\label{g_r_eq}\\
\frac{g_i}{g_r^2+g_i^2}&=\alpha\,g_i\int \frac{t^2}{(1-t\,g_r)^2+t^2 g_i^2}\,dP_*(t)\;,\label{g_i_eq}
\end{align}
where we denoted $dP_*(t):=p_*(t)\,dt$.
We remind that in the simple Gaussian case described in Eqs. (\ref{H}) and (\ref{GRF})  studied in \cite{fyodorov2004complexity}, the Hessian spectral density conditioned on stationarity retains its semi-circular form and is merely shifted by a $\mu$-dependent factor compared to its form at a generic position. In stark contrast, the conditioned Hessian spectral density obtained in this problem takes a different shape from the unconditioned one  (and thus cannot be recovered solely by a shift). It only coincides with the latter in the limit $\mu\to \infty$, where $p_*(t)=p_0(t)$.\\[0.5ex] 
To analyse further the ensuing  Hessian eigenvalue density $\rho(\lambda)$, we first consider the case where the bare density $p_0(t)$ has a finite left edge, inducing a similar edge for the optimal density $p_*(t)$. The conditioned Hessian spectrum $\rho(\lambda)$ similarly displays a finite left edge at a value $\lambda=\lambda_-$. This value together with $g_-\equiv m_r(\lambda_-;\alpha)$ can be determined by solving the following set of equations
\begin{align}
\frac{1}{g_{-}}&=\lambda_{-}-\alpha\int \frac{t}{1-t\,g_{-}}\,dP_*(t)\;,\\
\frac{1}{g_{-}^2}&=\alpha\,\int \frac{t^2}{(1-t\,g_{-})^2}\,dP_*(t)\;,
\end{align}
where $1/g_{-}$ lies outside the support of $p_*(t)$. Note that if the bare PDF $p_0(t)$ also has a finite right edge $\lambda_+$, its value as well as $g_+\equiv m_r(\lambda_+;\alpha)$ satisfies exactly the same set of equations. Solving the equations above, one may study the evolution of the quantity $\mu+\lambda_-$ ( positive in the trivial phase and negative in the complex phase) as a function of different parameters.  Analysing Eqs. (\ref{g_r_eq}-\ref{g_i_eq}) in the vicinity of the left edge, one finds that the conditioned Hessian spectral density retains the characteristic square-root edge behaviour:  
\be
\rho(\lambda)\approx \frac{\displaystyle |g_{-}|^{3/2}}{\displaystyle \pi\sqrt{1+\alpha\,\int dt\,p_*(t)\frac{t^3\,g_{-}^3}{(1-t\,g_{-})^3}}}\sqrt{\lambda-\lambda_-}\;,\;\;\lambda\to \lambda_-\;.
\ee

Next, considering a bare PDF $p_0(t)$ with unbounded support, the conditioned Hessian spectrum $\rho(\lambda)$ is similarly unbounded and we analyse its behaviour as $\lambda\to \pm \infty$. In that regime, $g_i\ll |g_r|\ll 1$ with $g_r\sim \lambda^{-1}$. Analysing Eq. \eqref{g_i_eq}, one can extract the leading density asymptotics as
\be
\rho(\lambda)\approx \alpha\,p_*(\lambda)\approx \alpha\,|\lambda|\sqrt{m_r^2+m_i^2}\,p_0(\lambda)\;,\;\;\lambda\to \pm\infty\;.
\ee

\subsection{Number of minima}

Let us now consider the problem of computing the expected number of minima for the random landscape described in Eq. (\ref{H}). The only difference with respect to the computation of the total number of critical points stems from conditioning the counting to the critical points with all stable directions, equivalent to restricting the lowest Hessian eigenvalue to be positive. The mean number of minima then reads
\begin{align}
    \mathbb{E}\left[{\cal N}_{\min}(\mu)\right]=\int d{\bf x}\,&\mathbb{E}\left[\prod_{i=1}^N\delta\left(\mu x_i+\sum_{l=1}^M k_{li}\phi_l'({\bf k}_l\cdot{\bf x})\right)\right.\\
    &\left.\times\det_{1\leq i,j\leq N}\left(\mu\delta_{ij}+\sum_{l=1}^M k_{li}k_{lj}\phi_l''({\bf k}_l\cdot{\bf x})\right)\Theta(\mu+\lambda_{min}(x))\right]\;,\nn
\end{align}
where $\lambda_{min}(x)$ stands for the lowest eigenvalue of the random matrix with entries $\sum_{l=1}^M k_{li}k_{lj}\phi_l''({\bf k}_l\cdot~{\bf x})$. Using the same steps as for evaluating the mean total number of critical points, one quickly arrives at the mean number of minima given  for any finite $N,M$ by
\begin{align}
\mathbb{E}\left[{\cal N}_{\min}(\mu)\right]&=\frac{1}{\mu^N}\mathbb{E}\left[\det\left(\mu\mathbb{I}+K T K^T\right)\Theta\left(\mu+\lambda_{min}\right)\right]\;,\label{n_min_res} 
\end{align}
where we denote $\lambda_{min}$ the lowest eigenvalue of the (position-independent) matrix $K T K^T$. The associated annealed complexity of minima is then given by
\be
\Sigma_{\min}(\mu,\alpha)=\lim_{N,M\to \infty}\frac{1}{N}\ln \mathbb{E}\left[{\cal N}_{\min}(\mu)\right]\;.
\ee
In full similarity with the computation of the total annealed complexity, we suppose that the strong self-averaging property described in Eq. \eqref{strong_self_av} holds conditioned on the fact that $\lambda_{min}>-\mu$. Applying again the machinery of the functional integral approach to large deviations one arrives at an extended optimisation problem which eventually controls the value of the annealed complexity of minima:
\begin{align}
 \Sigma_{\min}(\mu,\alpha)&=\max_{p(t):\int dt\,p(t)=1,w}S_{\alpha}[p(t),w]\label{sig_min_res}\\
 S_{\alpha}[p(t),w]&=\left[-\alpha\int p(t)\ln\left(\frac{p(t)}{p_0(t)}\right)\,dt+\int d\lambda\,\rho(\lambda)\ln\left|\frac{\mu+\lambda}{\mu}\right|\,d\lambda+w\int_{-\infty}^{-\mu} \rho(\lambda)\,d\lambda\right]\;,
 \end{align}
where the optimisation process is over the probability densities  $p(t)$ and the Lagrange parameter $w$ ensuring that the spectral density $\rho(\lambda)$ is now vanishing to the left of $-\mu$.
The stationarity equations read
\begin{align}
 \ln \frac{p_*(t)}{Z_p\,p_0(t)}&=\frac{1}{\alpha}\int d\lambda\left.\frac{\delta \rho(\lambda)}{\delta p(t)}\right|_{p(t)=p_*(t)}\ln\left|\frac{\mu+\lambda}{\mu}\right|+\frac{w}{\alpha}\int_{-\infty}^{-\mu}d\lambda \left.\frac{\delta \rho(\lambda)}{\delta p(t)}\right|_{p(t)=p_*(t)}\\
 &=\frac{1}{\alpha}\int_{\mu}^{\infty}d\nu\frac{\delta }{\delta p(t)}\left(m_r(-\nu;\alpha)+\frac{1}{\nu}+\frac{w}{\pi}m_i(-\nu;\alpha)\right)_{p(t)=p_*(t)}\nn\\
 &=\ln |1-t\,m(-\mu;\alpha)|+\frac{w}{\pi}{\rm arg}(1-t\,m(-\mu;\alpha))\;,  \nn
 \end{align}
 and
 \begin{align}
 \int_{-\infty}^{-\mu}\rho(\lambda)\,d\lambda&=\frac{1}{\pi}\int_{\mu}^{\infty} \,m_i(-\lambda;\alpha)\,d\lambda=0\;.
\end{align}
For these to hold, one needs to require that $m_i(-\lambda;\alpha)=0$ for any $\lambda\geq \mu$. As $m_i(-\lambda;\alpha)=\pi\rho(-\lambda)\geq 0$, the only possibility to satisfy this requirement is to ensure that the optimal density $p_*(t)$ has a finite left edge at $-t_m(\mu;\alpha)$ such that $p_*(t<-t_m)=0$. Once this is ensured, the remaining stationarity equation is exactly the same as for the total annealed complexity, with the optimal PDF 
\be
p_*(t)=\frac{\displaystyle p_0(t)(1-t\,m(-\mu;\alpha))\Theta(t+t_m(\mu,\alpha))}{\displaystyle\int_{-t_m(\mu,\alpha)}^{\infty} (1-t'\,m(-\mu;\alpha))p_0(t')\,dt'}\;.
\ee
Inserting such an optimal PDF in the equation \eqref{m_eq}, one obtains the following two equations for $t_m\equiv t_m(\mu;\alpha)$ and $m\equiv m(-\mu;\alpha)$,
\begin{align}
    \frac{1}{m}&=-\mu-\alpha\frac{\displaystyle\int_{-t_m}^{\infty}t\,dP_0(t)}{\displaystyle\int_{-t_m}^{\infty}(1-t\,m)\,dP_0(t)}\;,\label{m_r_eq_min}\\
    1&=\alpha \frac{\displaystyle\int_{-t_m}^{\infty}\frac{t^2\,m^2}{1-t\, m}\,dP_0(t)}{\displaystyle\int_{-t_m}^{\infty} (1-t\, m)\,dP_0(t)}\;,\label{m_i_eq_min}
\end{align}
with $t_m>0$ and $0>m>-1/t_m$. As in the case of the total annealed complexity, the annealed complexity of minima is expressed as
\be
\Sigma_{\min}(\mu,\alpha)=\int_{\mu}^{\infty}\left(m(-\nu,\alpha)+\frac{1}{\nu}\right)\,d\nu\;.
\ee
Considering a bare PDF $p_0(t)$ with a finite left edge $-t_{\rm e}$ and taking into account the fact that it has zero mean $\int_{-t_{\rm e}}^{\infty} t\,p_0(t)\,dt=0$, one can easily check from Eq. \eqref{m_r_eq_res_min} and the expression above that the complexity of minima vanishes above some threshold value $\mu_{\rm m}$, whose value is such that the imposed left edge $-t_{\rm m}$ coincides with the bare value $-t_{\rm e}$. At such a value $\mu_{\rm m}$ the Stieljes transform is given by $m=-\mu_{\rm m}^{-1}$. Inserting this in Eq. \eqref{m_i_eq_res_min} one recovers the same equation for the threshold value $\mu_{\rm m}$ as for the value $\mu_{\rm c}$. We conclude that the topology trivialisation transition for the total complexity and the complexity of minima thus occurs at the common threshold $\mu_{\rm m}=\mu_{\rm c}$. Conversely, for a bare PDF $p_0(t)$ with unbounded support, it is simple to check that $\int_{-t_{\rm m}}^{\infty} t\,p_0(t)\,dt\geq \int_{-\infty}^{\infty} t\,p_0(t)\,dt=0$ for any finite $t_{\rm m}$. For any finite $\nu$, Eq. \eqref{m_r_eq_res_min} thus shows that $m(-\nu,\alpha)+\frac{1}{\nu}>0$. The annealed complexity of minima is thus positive and monotonically decreasing for any finite value of $\mu$.

One may also analyse the behaviours of this annealed complexity of minima as $\mu \to 0$. It is simple to check that its leading asymptotic  is the same as for the total annealed complexity, i.e.
\be
\Sigma_{\min}(\mu,\alpha)\approx -\min(\alpha,1)\ln\mu\;,\;\;\mu\to 0\;.
\ee
Considering now the way that $\Sigma_{\min}(\mu,\alpha)$ vanishes, one needs to distinguish the case of a bare PDF $p_0(t)$ with a finite left edge (where the complexity vanishes at $\mu=\mu_c(\alpha)$) and the case of unbounded bare PDF where the complexity of minima only vanishes asymptotically as $\mu\to \infty$. To analyse the behaviour of the complexity, we find it useful to employ  the identity
\be
m(-\nu,\alpha)+\frac{1}{\nu}=-\frac{\alpha}{\mu}\frac{\displaystyle\int_{-t_m}^{\infty}m\,t\,dP_0(t)}{\displaystyle\int_{-t_m}^{\infty}(1-t\,m)\,dP_0(t)}=\frac{\alpha}{\mu}\frac{\displaystyle\int_{-\infty}^{-t_{m}}m\,t\,dP_0(t)}{\displaystyle 1-\int_{-\infty}^{-t_{\rm m}}(1-t\,m)\,dP_0(t)}\;.
\ee
For a bounded PDF, taking successive derivatives of  the complexity  close to $\mu_c(\alpha)$, one obtains at the first order
\be
\partial_\mu \Sigma_{\min}(\mu_c(\alpha),\alpha)=-\left(m(-\mu_c(\alpha);\alpha)+\frac{1}{\mu_c(\alpha)}\right)=-\frac{\alpha}{\mu_c(\alpha)}\frac{\displaystyle \int_{-t_{\rm e}}^{\infty}t\,dP_0(t)}{\displaystyle \int_{-t_{\rm e}}^{\infty}(\mu_c(\alpha)+t)\,dP_0(t)} =0\;.
\ee
The second order reads instead 
\begin{align}
 \partial_\mu^2 \Sigma_{\min}(\mu_c(\alpha),\alpha)&=-\frac{\alpha}{\mu_c(\alpha)^2}\partial_\mu\left(\int_{-t_{\rm m}}^{\infty}t\,dP_0(t) \right)_{\mu=\mu_c(\alpha)}\nn\\
 &=\frac{\alpha (\alpha I_2(\mu_c)-2\mu_c)(\mu_c-t_{\rm e})t_{\rm e}}{\mu_c^2\left[\mu_c^3+t_{\rm e}\left(\alpha^2 I_2(\mu_c)(\mu_c-t_{\rm e})+(1+\alpha)\mu_c(t_{\rm e}-2\mu_c)\right)\right]}\;.   
\end{align}
One may check that the expression above is positive, ensuring that the annealed complexity of minima vanishes quadratically at the topology trivialisation transition, see Eq. \eqref{min_edge}. As we have already mentioned, such a behaviour is quite distinct from the Gaussian toy model case  where the annealed complexity of minima vanished cubically at the transition \cite{fyodorov2007density,Fyodorov_Nadal_12},  see also \cite{fyodorov2020manifolds,BenArousBourgadeMcKenna24} for a similar behaviour for elastic manifold with finite internal dimension. 

For an unbounded PDF instead, one may use that $t_{\rm m}\sim \mu$ and $m\sim -\mu^{-1}$ as $\mu\to \infty$ to obtain that 
\be
m(-\mu,\alpha)+\frac{1}{\nu}\approx \frac{\alpha}{\mu^2}\int_{\mu}^{\infty}t\,p_0(-t)dt=\alpha\int_{1}^{\infty}dx\,x\,p_0(-\mu x)\;,\;\;\mu\to \infty\;.
\ee
The asymptotic behaviour of $\Sigma_{\min}(\mu,\alpha)$ as $\mu\to \infty$ 
is thus the same as for the total annealed complexity $\Sigma_{\rm tot}(\mu,\alpha)$ up to a multiplicative factor $1/2$.

\vspace{0.3cm}

{\it Some properties of the spectral density for Hessian at points of local minima}

Let us conclude this section by mentioning some aspects of the Hessian spectral density $\rho(\lambda)$  arising from this minimisation procedure for the complexity of minima. It may be studied by considering the behaviour of the Stieltjes transform at position $\lambda\in \mathbb{R}$. Keeping the same notation $m\equiv m(-\mu,\alpha)$ and $t_{\rm m}\equiv t_{\rm m}(\mu,\alpha)$ and introducing $g_r\equiv m_r(\lambda,\alpha)$ and $g_i\equiv m_i(\lambda,\alpha)$, the latter satisfy the following set of equations
\begin{align}
\frac{g_r}{g_r^2+g_i^2}&=\lambda-\alpha\frac{\displaystyle\int_{-t_{\rm m}}^{\infty} (1-t\,m)\,\frac{t\,(1-t\,g_r)}{(1-t\,g_r)^2+t^2 g_i^2}dP_0(t)}{\displaystyle\int_{-t_{\rm m}}^{\infty} (1-t\,m)\,dP_0(t)}\;,\label{g_r_eq_1}\\
\frac{g_i}{g_r^2+g_i^2}&=\alpha\,g_i\frac{\displaystyle\int_{-t_{\rm m}}^{\infty} (1-t\,m)\frac{t^2}{(1-t\,g_r)^2+t^2 g_i^2}\,dP_0(t)}{\displaystyle\int_{-t_{\rm m}}^{\infty} (1-t\,m)\,dP_0(t)}\;.\label{g_i_eq_1}
\end{align}
One may first consider positions $\lambda\leq -\mu$, where the Hessian spectral density should be identically zero, i.e. $g_i=\pi\rho(\lambda)=0$. In that case the corresponding value of $g_r$ satisfies the following simpler equation
\be
\frac{1}{g_r}=\lambda-\alpha\frac{\displaystyle\int_{-t_{\rm m}}^{\infty} \frac{t\,(1-t\,m)}{1-t\,g_r}\,dP_0(t)}{\displaystyle\int_{-t_{\rm m}}^{\infty} (1-t\,m)\,dP_0(t)}\;,
\ee
which requires that $g_r>-t_{\rm m}^{-1}$. One may compute the derivative of $g_r$ with respect to $\lambda$, yielding
\begin{align}
\frac{1}{g_r'}&=\frac{1}{g_r^2}\left(-1+\alpha \frac{\displaystyle\int_{-t_m}^{\infty}\frac{t^2\,g_r^2(1-t m)}{(1-t\, g_{r})^2}\,dP_0(t)}{\displaystyle\int_{-t_m}^{\infty} (1-t\, m_{r})\,dP_0(t)}\right)\\
&=-\frac{2\alpha m(g_r-m)}{ \displaystyle g_r\int_{-t_m}^{\infty} (1-t\, m_{r})dP_0(t)}\int_{-t_m}^{\infty}\frac{t^2}{(1-t\, g_{r})^2(1-t\,m)}\left[t-\frac{1}{2}\left(\frac{1}{g_r}+\frac{1}{m}\right)\right]dP_0(t)\;.
\end{align}
Using that both $g_r>-1/t_{\rm m}$ and $m>-1/t_{\rm m}$, it is simple to check that this expression is of the opposite sign with the difference $g_r-m$. We therefore conclude that $g_r>m$ and it is a monotonically decreasing function behaving as $g_r\sim 1/\lambda$ for $\lambda\to -\infty$ and $g_r\to m$ for $\lambda\to -\mu$.

Let us now consider the behaviour for $\rho(\lambda)$ at $\lambda>-\mu$. For any bare density $p_0(t)$ the Hessian spectrum associated to counting minima displays a finite left edge at a certain $\lambda_-$, which may differ from the value $-\mu$ but must satisfy $\lambda_-+\mu\geq 0$. To compute its value, one needs to solve the following equations
\begin{align}
  \frac{1}{g_-}&=\lambda_- -\alpha\frac{\displaystyle\int_{-t_{\rm m}}^{\infty}\frac{t\,(1-t\,m)}{1-t\,g_-}\, dP_0(t)}{\displaystyle\int_{-t_{\rm m}}^{\infty} (1-t\,m)\, dP_0(t)}\;,\\
\frac{1}{g_-^2}&=\alpha\,\frac{\displaystyle\int_{-t_{\rm m}}^{\infty} \frac{t^2\,(1-t\,m)}{(1-t\,g_-)^2}\, dP_0(t)}{\displaystyle\int_{-t_{\rm m}}^{\infty} (1-t\,m)\, dP_0(t)}\;, 
\end{align}
where we have denoted $g_-\equiv m_r(\lambda_-;\alpha)$. In the complex phase we then find the value $\lambda_-=-\mu$ and the gap $\delta=\lambda_-+\mu=0$. For a PDF $p_0(t)$ with a finite left edge there exists a trivial phase where the complexity is zero, with $t_{\rm m}=t_{\rm e}$ and $m=-\mu^{-1}$ such that the gap $\delta=\lambda_-+\mu>0$ is positive. Its expression can be obtained by solving the equations above. The Hessian spectral density can be expanded in the vicinity of the left edge $\lambda_-$, displaying the following square-root behaviour
\be
\rho(\lambda)\approx \frac{\displaystyle |g_{-}|^{3/2}}{\displaystyle \pi\sqrt{1+\alpha\,\int_{-t_{\rm m}}^{\infty} \frac{t^3\,g_{-}^3\,(1-t\,m)}{(1-t\,g_{-})^3}}\,dP_0(t)}\sqrt{\lambda-\lambda_-}\;,\;\;\lambda\to \lambda_-\;.\label{rho_sq_ro}
\ee
For a PDF $p_0(t)$ with a finite right edge as well, the Hessian spectrum $\rho(\lambda)$ will have a finite right edge $\lambda_+$, satisfying together with the Stietjes transform at this value $g_+\equiv m_r(\lambda_+;\alpha)$ the same set of equation as $\lambda_-$ and $g_-$. The behaviour in its vicinity is again of square-root type as in Eq. \eqref{rho_sq_ro}. Finally, if the PDF $p_0(t)$ does not have any finite right edge, the asymptotic behaviour of the Hessian spectrum as $\lambda\to +\infty$ reads
\be
\rho(\lambda)\approx \alpha |m|\lambda\,p_0(\lambda)\;,\;\;\lambda\to +\infty\;.
\ee

\section{Optimisation problem for the ground-state energy} \label{sec_opt_prob}

In this section, we derive the expression for the mean ground-state energy $e_0$ in terms of the optimisation problem given in Eq. \eqref{gen_opt}. Our starting point is the expression for the positive integer moments of the partition function evaluated at $N\to \infty$, from which we will be able to express the mean free energy using the replica trick. Finally, we will consider the zero-temperature limit to obtain our final expression for the mean ground-state energy.

\subsection{Integer moments of the partition function}

We start with defining the partition function associated with our landscape as
\be\label{partfun}
{\cal Z}(\beta)=\int_{\mathbb{R}^N} e^{-\beta {\cal H}({\bf x})}\, d{\bf x},
\ee
where $\beta>0$ stands for the inverse temperature.
Next we follow the paradigm of the replica trick and compute the positive integer moments $\mathbb{E}\left[{\cal Z}(\beta)^{n}\right]$ for $n\in\mathbb{N}$:
\begin{align}
    \mathbb{E}\left[{\cal Z}(\beta)^{n}\right]=\int_{\mathbb{R}^N} \mathbb{E}\left[e^{-\beta\sum_{a=1}^n {\cal H}({\bf x}_a)}\right]\,\prod_{a=1}^n d{\bf x}_a\,&\nn \\ =\int_{\mathbb{R}^N} e^{-\frac{\beta\mu}{2}\sum_{a=1}^n {\bf x}_a^2}\,\prod_{l=1}^M\mathbb{E}\left[e^{-\beta\sum_{a=1}^n \phi_l({\bf k}_l\cdot{\bf x}_a)}\right]\prod_{a=1}^n d{\bf x}_a
   & =\int_{\mathbb{R}^N} e^{-\frac{\beta\mu}{2}\sum_{a=1}^n {\bf x}_a^2}\,\left(V_n\right)^M\,\prod_{a=1}^n d{\bf x}_a\;,\label{replicated}
\end{align}
where we used independence of $\phi_l(z)$ for different $l=1,\ldots,M$ and defined
\be
V_n\equiv V_n(\{{\bf x}_a\,,\,\,a=1,\cdots,n\})=\mathbb{E}_{\bf \phi}\left[\mathbb{E}_{\bf k}\left[e^{-\beta\sum_{a=1}^n \phi({\bf k}\cdot{\bf x}_a)}\right]\right]\;.\label{V_n}
\ee
We can further write 
\be \label{expectation}
\mathbb{E}_{\bf k}\left[e^{-\beta\sum_{a=1}^n \phi({\bf k}\cdot{\bf x}_a)}\right]=
\int e^{-\beta\sum_{a=1}^n \phi(z_a)} {\cal P}(z_1,\ldots,z_n)\,\prod_{a=1}^n dz_a
\ee
where we defined the joint probability density 
\be
{\cal P}(z_1,\ldots,z_n)=\mathbb{E}_{\bf k} \left[\prod_{a=1}^n \delta\left(z_a-{\bf k}\cdot{\bf x}_a\right)\right],
\ee
with the latter being further represented via the Fourier-transform as
\be \label{joint_prob_z}
{\cal P}(z_1,\ldots,z_n)=\int 
 e^{i\sum_{a=1}^n u_az_a}\mathbb{E}_{\bf k}\left[e^{-i{\bf k}\cdot{\bf b}}\right] \prod_{a=1}^n \frac{du_a}{2\pi}, \quad {\bf b}=\sum_{a=1}^nu_a{\bf x}_a.
\ee
Now we may have two different choices for the statistics of random wavevectors ${\bf k}$.

\begin{enumerate}
    \item When the vector ${\bf k}$ is a $N$-dimensional vector of i.i.d. Gaussian components, each with zero mean and variance $1/N$, we immediately have 
    \be\label{Gauav}
    \mathbb{E}_{\bf k}\left[e^{-i{\bf k}\cdot{\bf b}}\right]=e^{-\frac{1}{2N}{\bf b}^2}=e^{-\frac{1}{2}\sum_{a,b}u_au_b\,\left(\frac{{\bf x}_a\cdot {\bf x}_b}{N}\right)}
    \ee

    \item When the wavevector ${\bf k}$ is drawn uniformly on the $N$-sphere $\sqrt{N}{\cal S}_{N-1}$, we can write 
     \be\label{sphere}
    \mathbb{E}_{\bf k}\left[e^{-i{\bf k}\cdot{\bf b}}\right]=
    C_N\int e^{-i{\bf k}\cdot{\bf b}}\delta({\bf k}^2-1)d{\bf k}, 
    \ee
    where $C_N$ is an appropriate normalization constant.  Passing to the polar coordinates allows one to reduce the above integral to 
    \be
    C_N\int e^{-i{\bf k}\cdot{\bf b}}\delta({\bf k}^2-1) d{\bf k}=\frac{\Gamma\left(\frac{N}{2}\right)}
    {\Gamma\left(\frac{N-1}{2}\right)\Gamma\left(\frac{1}{2}\right)}\int_0^{\pi}e^{i|{\bf b}|\cos{\theta}}\sin^{N-2}\theta\,d\theta
    \ee
    and further changing to $t=\sqrt{N}\cos{\theta}$ we see that 
    \be
     \mathbb{E}_{\bf k}\left[e^{-i{\bf k}\cdot{\bf b}}\right]=\frac{\Gamma\left(\frac{N}{2}\right)}
    {\sqrt{N}\Gamma\left(\frac{N-1}{2}\right)\Gamma\left(\frac{1}{2}\right)}\int_{-\sqrt{N}}^{\sqrt{N}}e^{it
    \frac{|{\bf b}|}{\sqrt{N}}}\left(1-\frac{t^2}{N}\right)^{\frac{N-3}{2}}\,dt.
    \ee
The most relevant limit to us corresponds to $N\to \infty$ keeping $|{\bf b}|/\sqrt{N}$ finite. In such a limit the integral over $t$ becomes Gaussian and reproduces exactly the formula Eq. \eqref{Gauav}. Hence both types of models for random wave vectors ${\bf k}$ become asymptotically equivalent as $N\to \infty$.  

\end{enumerate}  

Now  substituting  Eq. \eqref{Gauav} to Eq. \eqref{joint_prob_z} the integral over the variables $u_a$ becomes Gaussian and 
can be readily evaluated, yielding the Gaussian joint probability density for variables $z_a, \, a=1,\ldots n$ represented as a (column) vector ${\bf z}=(z_1,\ldots,z_n)^T$:
\be \label{jpdz}
{\cal P}(z_1,\ldots,z_n)=\frac{1}{(2\pi)^{n/2}\sqrt{\det{R}}}\exp\left\{-\frac{1}{2}{\bf z}^TR^{-1}{\bf z}\right\}, \quad 
R_{ab}=\frac{{\bf x}_a\cdot{\bf x}_b}{N}
\ee

 Substituting Eq. \eqref{jpdz} into \eqref{expectation} we finally arrive to the expression of $V_n$ in Eq. \eqref{V_n} in the form
\begin{align}
   V_n=\frac{\sqrt{\det Q}}{(2\pi)^{n/2}}\, W_n(Q), \quad W_n(Q)=\mathbb{E}_{\phi}\left[\int e^{-\frac{1}{2}{\bf z}^T Q{\bf z}}\,e^{-\beta\sum_{a=1}^n \phi(z_a)}\, \prod_{a=1}^n\,dz_a\,\right]\;,\;\;Q=R^{-1}\;.\label{W_n}
\end{align}
We thus see that the integrand in Eq. \eqref{replicated} depends on the vectors ${\bf x}_a$ only via the matrix of their scalar products $R_{ab}$.
This allows to apply the method of invariant integration developed in \cite{fyodorov2007classical} and replace 
the integral over ${\bf x}_a$ in \eqref{replicated} first with the integral over positive definite $n\times n$ overlap matrices $R$,  incurring in the process the Jacobian 
$\left(\det{R}\right)^{\frac{N-n-1}{2}}$. However in contrast to by now standard computations in Gaussian models (either the toy model \eqref{GRF} of elastic manifolds or a spherical spin glass model) we found it to be more convenient in the present case to use the {\it inverse} of the matrix $R$ of replica overlaps as the integration variable. Passing from $R$ to
 $Q=R^{-1}$ incurrs further Jacobian factor $(\det{Q})^{n-1}$ and one finally arrives at representing the  moments of the partition function for any finite integer value of $n$ satisfying $n(n+1)/2<N$ as
\begin{align}
    \mathbb{E}\left[{\cal Z}(\beta)^{n}\right]&=c_{N,n}\int_{Q>0}\, (\det Q)^{-\frac{n+1}{2}}e^{N\,\Sigma_{n}(Q)}\,dQ\,,
 \end{align}
 with the known constants $c_{N,n}$ and the terms in the exponential given by
 \begin{align}
    \Sigma_{n}(Q)&=-\frac{\beta\mu}{2}\Tr Q^{-1}+\frac{\alpha-1}{2}\ln \det Q+\alpha\ln W_n(Q)+\frac{n}{2}\left[(1-\alpha)\ln(2\pi)+1\right]\;.\label{action}
\end{align}
 In order to obtain the mean ground-state energy as $N\to \infty$ one may now employ the Laplace method which requires finding stationarity point of $\Sigma_{n}(Q)$ for finite $n$ and $\beta$  and then eventually considering  the limit $n\to 0$ extracting the mean free-energy of the model. Finally taking the limit $\beta\to\infty$ yields the mean ground state. In pursuing this way we follow the standard techniques of the replica method  by introducing the Parisi function, see e.g. \cite{parisi_urbani_zamponi2020book}.

\subsection{Introduction of the Parisi function and the replica trick}

 Following the standard assumptions of the Parisi method we postulate that the (inverse) matrix of overlaps $Q$ satisfying the stationarity of  $\Sigma_{n}(Q)$ in Eq. \eqref{action} takes a special block form corresponding to an arbitrary number $k$ of the levels of replica symmetry breaking. The associated sizes of the blocks describing this structure for the matrix $Q$  for an integer $n$ take values
\be
m_0=n\geq m_1\geq \cdots\geq m_k\geq m_{k+1}=1\;,\label{m_ineq}
\ee
while the matrix entries within the corresponding blocks take values
\be
0>q_0\geq q_1\geq \cdots\geq q_k\ \quad \mbox{and} \quad  q_d>0,
\ee
with $q_d$ standing on the principal diagonal.
It will be convenient to work in the following with the matrix $P=(q_d-q_k)\mathbb{I}+q_0{\bf 1}-Q$, where the matrix elements are $\mathbb{I}_{ab}=\delta_{ab}$ and ${\bf 1}_{ab}=1, \, \forall (a,b)$. This matrix $P$ has the same block sizes structure as $Q$ and its entries taking the values 
\be
p_0=0\leq p_1=q_0-q_1\leq \cdots\leq p_i=q_0-q_i\leq \cdots\leq p_k=r_d=q_0-q_k\;.
\ee
with blocks being ordered from the largest possible size to the smallest size $1$ on the diagonal. 
Next we introduce the Parisi function $x(p)$, helping to parametrize the matrix $P$, via
\be\label{stepX}
x(p)=n+\sum_{j=0}^{k}(m_{j+1}-m_{j})\Theta(p-p_{j})\;,
\ee
with $\Theta(x)$ standing for the Heaviside step function. Note that the Parisi function defined here describes rescaled overlaps only in the interval $[0,p_k]$. Such a form implies, in particularly,  
\be\label{stepinvX}
\frac{1}{x(p)}=\frac{1}{n}+\sum_{j=0}^{k}\left(\frac{1}{m_{j+1}}-\frac{1}{m_{j}}\right)\Theta(p-p_{j})\;.
\ee
We further introduce the following definitions
\be\label{defin}
\nu=\frac{q_d-q_k}{\beta}\quad ,\quad \lambda(p)=\beta\nu -\int_{p}^{{p_k}}x(\tilde{p})\,d\tilde{p}\;,
\ee
which will be used throughout the manuscript to express the eigenvalues of the matrix $Q$. Note that at this stage both the functions $x(p)$ and $\lambda(p)$ are piece-wise constant. The matrix $Q$ has $k+2$ distinct eigenvalues $\lambda_i$'s of degeneracies $d_i$'s given respectively by
\begin{align}
\lambda_{k+1}&=\beta\nu=q_d-q_k\;,\;\;d_{k+1}=n\left(1-\frac{1}{m_{k}}\right)\;,\\
\lambda_i=&\lambda(p)\;,\;\;p_{i-1}<p<p_i\;,\;\;d_i=n\left(\frac{1}{m_{i}}-\frac{1}{m_{i-1}}\right)\;,\;\;i=1,\cdots,k+1\;,\nn\\
\lambda_0=&\lambda_1+n\,q_0\;,\;\;d_0=1\;.
\end{align}
Using these formulae one may express explicitly the trace of any function $G(Q)$ in the following form:
\begin{align}
   & \Tr\left[G(Q)\right]=G(\lambda_0)-G(\lambda_1)+n\,G(\lambda_{k+1})-\frac{n}{m_k}\,\left[G(\lambda_{k+1})-G(\lambda_1)\right]  \nn \\& +n\sum_{i=1}^{k}\left(\frac{1}{m_i}-\frac{1}{m_{i-1}}\right)\,\left[G(\lambda_i)-G(\lambda_1)\right] 
    \end{align}
which using  Eq. \eqref{stepinvX} and Eq. \eqref{defin} can be further rewritten as  
    \begin{align}  \nn
    \Tr\left[G(Q)\right]  =G\left(\lambda(0)+n q_0\right)-G\left(\lambda(0)\right)+ & n\,G(\beta\nu)-n\int_0^{p_k}\left[G(\lambda(p))-G(\lambda(0))\right]\, \frac{x'(p) }{x(p)^2}\,dp\,\\
    =G\left(\lambda(0)+n q_0\right)-G\left(\lambda(0)\right)+& n\,G(\beta\nu)-  n\int_0^{p_k}G'(\lambda(p))\,dp\;.
\end{align}
At the next step we aim at expressing the term $W_n$ in Eq. \eqref{W_n}  in terms of the  matrix $P$ and its associated Parisi function. We start with the following representation: 
\be
W_n=\mathbb{E}_{\phi}\left[\int d{\bf z}\,e^{\frac{1}{2}\sum_{a,b=1}^n P_{ab}z_a z_b-\beta\sum_{a=1}^n\left[\frac{\nu}{2}z_a^2+\phi(z_a)\right]-\frac{q_0}{2}\left(\sum_{a=1}^n z_a\right)^2}\right]\;,
\ee
and apply the identity first introduced in this context by Duplantier \cite{D81} and used extensively thereafter \cite{parisi_urbani_zamponi2020book}
\be
\exp\left(\frac{1}{2}\sum_{a,b=1}^n P_{ab}z_a z_b\right)=\lim_{{\bf v}\to {\bf 0}}\exp\left(\frac{1}{2}\sum_{a,b=1}^n P_{ab}\partial_{v_a}\partial_{v_b}+{\bf v}\cdot {\bf z}\right)\;.
\ee
This identity can be used to recast $W_n$ as
\begin{align}
 W_n=\mathbb{E}_{\phi}\left[\lim_{{\bf v}\to {\bf 0}}\exp\left(-\frac{q_0}{2}\left(\sum_{a}\partial_{v_a}\right)^2+\frac{1}{2}\sum_{a,b=1}^n P_{ab}\partial_{v_a}\partial_{v_b}\right)\int d{\bf z}\,e^{-\beta\sum_{a=1}^n\left[\frac{\nu}{2}z_a^2-\frac{v_a z_a}{\beta}+\phi(z_a)\right]}\right] \nn
 \end{align}
 \begin{align}
 =\mathbb{E}_{\phi}\left[\int \frac{dv}{\sqrt{-2\pi q_0}}e^{\frac{v^2}{2 q_0}+n g_n(0,v)}\right]\;, 
\end{align}
where the function $g_n(p,v)$ satisfies the partial differential equation
\be
\partial_p g_n=-\frac{1}{2}\left[\partial_{v}^2 g_n+x(p)(\partial_{v} g_n)^2\right]\;,\;\;g_n(p>p_k,v)=\ln\int e^{-\beta H_{\nu,v/\beta}(z)}\,dz\;,\label{Par_eq}
\ee
with the one-dimensional disordered Hamiltonian $H_{\nu,h}(z)$ being introduced earlier in Eq. \eqref{H_e_min}. Next, we can use the stationarity of the random function $\phi(z)$, to simplify the expression above as (see Appendix \ref{app_q0} for details)
\begin{align}
  \mathbb{E}_{\phi}\left[\int \frac{dv}{\sqrt{-2\pi  q_0}}e^{\frac{v^2}{2 q_0}+n g_n(0,v)}\right]&=\int e^{\frac{v^2}{2 q_0}}\mathbb{E}_{\phi}\left[e^{n g_n(0,v)}\right]\,\frac{dv}{\sqrt{-2\pi q_0}}\nn \\=\int e^{\frac{v^2}{2 }\left(\frac{1}{q_0}+\frac{n}{\lambda(0)}\right)}\,\frac{dv}{\sqrt{-2\pi q_0}}\,\mathbb{E}_{\phi}\left[e^{n g_n(0,0)}\right]
  &=\left(1+\frac{n q_0}{\lambda(0)}\right)^{-\frac{1}{2}}\mathbb{E}_{\phi}\left[e^{n g_n(0,0)}\right]\;.  \label{removing_q0}
\end{align}
Finally, collecting all the terms, we obtain the following expression for the function in Eq. \eqref{action}:
\begin{align}
    \Sigma_{n}(Q)=&-\frac{\beta\mu}{2}\left[\frac{1}{\lambda(0)+n q_0}-\frac{1}{\lambda(0)}+\frac{n}{\beta\nu}+n\int_0^{p_k}\frac{dp}{\lambda(p)^2}\right] \nn\\ &-\frac{1-\alpha}{2}\left[\ln\left(1+\frac{n q_0}{\lambda(0)}\right)+  n\ln(\beta\nu)-n\int_0^{p_k}\frac{dp}{\lambda(p)} \right] \\
    &+\alpha\ln \mathbb{E}_{\phi}\left[e^{n g_n(0,0)}\right]-\frac{\alpha}{2}\ln\left(1+\frac{n q_0}{\lambda(0)}\right)+\frac{n}{2}\left[(1-\alpha)\ln(2\pi)+1\right]\;\nn.
    \end{align}
The final step in obtaining the free-energy variational functional requires taking the appropriate limit $n\to 0$ of the above expression, getting
\begin{align} \label{Fre_energy_variational}
 {\cal F}_0[x(p)]=-\lim_{n\to 0}\frac{\Sigma_{n}(Q)}{n\beta}
 =\frac{q_0}{2\lambda(0)}\left(\frac{\mu}{\lambda(0)}-1\right) & \\+\frac{\mu}{2}\left[\frac{1}{\beta\nu}-\frac{1}{\beta\mu}+\int_0^{p_{m}}\frac{dp}{\lambda(p)^2}\right] & +\frac{1-\alpha}{2\beta}\left[\ln\frac{\beta\nu}{2\pi}-\int_0^{p_{m}}\frac{dp}{\lambda(p)}\right]-\frac{\alpha}{\beta}\Ep{g(0,0)}\;\nn
\end{align}
where we have denoted $p_m=\lim_{n\to 0}p_k$ so that $\lambda(p)=\beta\nu -\int_{p}^{{p_m}}x(\tilde{p})\,d\tilde{p}$. optimising this functional with respect to $x(p)$, $p_m$, $\nu$ and $q_0$, one can compute the mean free-energy of our model at a finite temperature $0<\beta<\infty$. In particular, by considering the zero-temperature limit $\beta\to \infty$ of this object results in a variational problem for the mean ground-state energy $e_0$.

\subsection{Zero-temperature scaling}

To extract the mean ground-state energy, one needs to introduce proper rescaling of the parameters and Parisi functions with $\beta$ allowing to perform the zero-temperature limit.   In particular, one anticipates the following parameters to remain finite and positive at 
$\beta\to \infty$:
\be
\nu=\lim_{\beta\to\infty}\frac{q_d-q_k}{\beta}\;,\;\;l=\lim_{\beta\to\infty}\frac{q_0-q_k}{\beta^2}=\lim_{\beta\to\infty}\frac{p_m}{\beta^2}\;,\;\;\;l_0=-\lim_{\beta\to\infty}\frac{q_0}{\beta^2}.
\ee

Accordingly, one rescales the Parisi function $x(p)$, as well as the eigenvalue function $\lambda(p)$, and  defines the following functions
\be \label{w_and_lambda}
w(t)=\lim_{\beta\to\infty}\beta x(\beta^2 t)\;,\;\;\lambda_0(t)=\lim_{\beta\to\infty}\frac{\lambda(\beta^2 t)}{\beta}=\nu-\int_t^{l}d\tau\,w(\tau)\;,\;\;t\in[0,l]\;.
\ee
Finally, one needs to consider the associated  rescaling of the Parisi differential equation \eqref{Par_eq}. One may naturally assume the following limit to exist as $\beta\to \infty$:
\be
f(t,h)=\lim_{\beta\to \infty}\frac{g_n(\beta^2 t,\beta h)}{\beta}\;,\;\;t\in[0,l]\;,\;\;h\in \mathbb{R}\;,
\ee
where the function $f(t,h)$ satisfies the Parisi partial differential equation
\be
\partial_t f=-\frac{1}{2}\left[\partial_{h}^2 f+w(t)(\partial_{h} f)^2\right]\;\;,\label{Parisi_eqA}
\ee
with a well-defined boundary condition, corresponding to the ground-state energy of the effective one-dimensional disordered Hamiltonian $H_{\nu,h}(z)$ in \ref{H_e_min}:
\be f(t\geq l,h)=-\epsilon_{\min}(\nu,h)\;.\label{Parisi_eq_bc}
\ee
It is worth noting that the zero-temperature limit in the present model seems more benign
than in the paradigmatic Sherrington-Kirkpatrick model, where the boundary condition for the Parisi equation reads at finite $\beta$ 
\be
g_n(p\geq p_k,v)=\frac{1}{\beta}\ln\cosh(v)\;,
\ee
and is thus non-analytic in the limit $\beta\to \infty$, i.e. 
\be
f(p,h)=\lim_{\beta\to \infty}\frac{g_n(p,\beta h)}{\beta}\;,\;\;f(p\geq p_{m},h)=|h|\;.
\ee
In contrast, in the present model the function $\epsilon_{\min}(\nu,h)$ is an analytic function of $h$. 

Inserting these different scaling forms in the expression \eqref{Fre_energy_variational} and taking the limit $\beta\to \infty$, the ground-state energy is finally expressed as
\begin{align}
   e_0=\underset{l,\nu,l_0,w(l')}{\rm sup}\left[\frac{l_0}{2\lambda_0(0)}\left(\frac{\mu}{\lambda_0(0)}-1\right)+\frac{\mu}{2}\int_{0}^{l}\frac{dt}{\lambda_0(t)^2}-\frac{1-\alpha}{2}\int_{0}^{l}\frac{dt}{\lambda_0(t)}
    -\alpha\mathbb{E}_{\phi}\left[f(0,0)\right]\right]\;.\label{GS_opt_prob}
\end{align}
The optimisation with respect to the parameter $l_0$ is rather trivial and yields $\lambda_0(0)=\mu$, which using Eq. \eqref{w_and_lambda} may be rewritten as
\be
\nu=\mu+\int_0^l w(t)\,dt\;.\label{nu_eq}
\ee
The stationarity condition with respect to $\nu$ allows to obtain an explicit expression for $l_0$  which reads
\be
\frac{l_0}{\mu^2}=(1-\alpha)\int_{0}^{l}\frac{dt}{\lambda_0(t)^2}-2\mu\int_{0}^{l}\frac{dt}{\lambda_0(t)^3}-2\alpha\partial_{\nu}\mathbb{E}_{\phi}\left[f(0,0)\right]\;.\label{l_0_eq}
\ee
Before analysing in more details our optimisation problem, let us introduce an alternative expression for the average ground-state energy that proves to be more convenient in our analysis.  In doing so we follow the method proposed in  \cite{parisi_urbani_zamponi2020book} and  introduce a functional Lagrange parameter $P(t,h)$, which ensures that the function $f(t,h)$ satisfies the Parisi differential equation with the correct associated boundary condition. Adding the corresponding terms to the optimisation functional yields:
\begin{align}
   e_0=&\underset{l,\nu,l_0,w(l),f(t,h),P(t,h)}{\rm sup}\left\{\frac{l_0}{2\lambda_0(0)}\left(\frac{\mu}{\lambda_0(0)}-1\right)+\frac{\mu}{2}\int_{0}^{l}\frac{dt}{\lambda_0(t)^2}-\frac{1-\alpha}{2}\int_{0}^{l}\frac{dt}{\lambda_0(t)}\right.\label{GS_opt_prob_2}\\
   &-\alpha\mathbb{E}_{\phi}\left[\int f(0,h)\delta(h)\,dh\right]+\alpha\Ep{\int P(l,h)\left(f(l,h)+\epsilon_{\min}(\nu,h)\right)\,dh}\nn\\
   &\left.-\alpha\Ep{\int_0^l dt\,\int dh\,P(t,h)\left(\partial_t f+\frac{1}{2}\left[\partial_{h}^2 f+w(t)(\partial_{h} f)^2\right]\right)}\right\}\;.\nn
\end{align}
In addition to the stationarity equations obtained from the previous form of the optimisation functional, one now also may show that $P(t,h)$ must satisfy the following partial differential equation and the boundary condition:
\be
\partial_t P=\frac{1}{2}\partial_{h}\left[\partial_{h}-2w(t)\partial_{h}f\right]P\;,\;\;P(0,h)=\delta(h)\;.\label{Parisi_eqB}
\ee
In addition, comparing the stationarity condition over the parameter $\nu$ obtained from Eq. \eqref{GS_opt_prob_2} with Eq. \eqref{l_0_eq}, one obtains a somewhat more explicit expression for the quantity
\be
\partial_{\nu}\mathbb{E}_{\phi}\left[f(0,0)\right]=-\Ep{\int\partial_{\nu}\epsilon_{\min}(\nu,h)\, 
 P(l,h) dh}\;.
\ee
With both expressions \eqref{GS_opt_prob} and \eqref{GS_opt_prob_2} being at our disposal for determining the ground-state energy $e_0$ in terms of the optimisation problems, we are now in a position to analyse the ensuing expression for the ground-state energy in different phases.

\section{Analysis of the different phases of the ground-state energy} \label{sec_analysis}

In this section, we will analyse in detail the different phases of the ground-state energy.

\subsection{Replica symmetric phase}

In the replica symmetric phase, the matrix $Q$ reads
\be
Q_{ab}=\begin{cases}
    q_d&\;,\;\;a=b\;,\\
    q_0&\;,\;\;a\neq b\;.
\end{cases}
\ee
The matrix $P=(q_d-q_0)\mathbb{I}+q_0{\bf 1}-Q$ is the null matrix in the replica symmetric (RS) phase, implying that the associated rescaled Parisi function is identically zero $w(t)=0$, which can be ensured by taking the optimal value $l=0$. Under these conditions the optimisation problem in Eq. \eqref{GS_opt_prob} simplifies dramatically and reads 
\be
e_{\rm RS}=\underset{\nu,l_0}{\rm max}\left[\frac{l_0}{2\nu}\left(\frac{\mu}{\nu}-1\right)+\alpha\mathbb{E}_{\phi}\left[\epsilon_{\min}(\nu)\right]\right]=\alpha \mathbb{E}_{\phi}\left[\epsilon_{\min}(\mu)\right]\;,
\ee
where we have used Eqs. \eqref{l_0_eq} and \eqref{nu_eq} yielding the following optimal value of $l_0$ and $\nu$:
\be
\nu=\mu\;,\;\;l_0=\alpha \mu^2\mathbb{E}_{\phi}\left[z_{\min}(\mu)^2\right]\;.
\ee
In the limit $n\to 0$, the matrix $R$ of replica overlaps takes the form
\be
R_{ab}=\frac{{\bf x}_a\cdot{\bf x}_b}{N}=\begin{cases}
    \displaystyle r_d=\frac{1}{q_d-q_0}\left(1-\frac{q_0}{q_d-q_0}\right)&\;,\;\;a=b\;,\\
    &\\
    \displaystyle r=-\frac{q_0}{q_d-q_0}&\;,\;\;a\neq b\;,
\end{cases}
\ee
from which one can compute the expression of the (rescaled) overlaps in the zero-temperature limit
\be
v=\lim_{\beta\to \infty}\beta(r_d-r)=\frac{1}{\mu}\;,\;\;r_0=\lim_{\beta\to \infty}r=\lim_{\beta\to \infty}r_d=\frac{l_0}{\mu^2}=\alpha\Ep{z_{\min}^2(\mu)}\;.
\ee
Let us now consider the condition under which this solution becomes unstable and one transits into a replica symmetry broken phase.

\subsection{Continuous replica symmetry breaking transition}

The most standard method to test the stability of the RS ansatz amounts to expanding the function $\Sigma_n(Q)$ in Eq. \eqref{action} up to second order around the RS solution and to compute the eigenvalues of the quadratic form obtained from that expansion \cite{AT78}. For the solution to be stable, one must ensure that the {\it replicon} eigenvalue, i.e. the largest eigenvalue of this quadratic form, is negative. This is quite a tedious computation whose details can be found in the Appendix \ref{replicon}. Following  \cite{parisi_urbani_zamponi2020book} we describe below an alternative method allowing not only to provide the location of the de Almeida-Thouless line separating the RS and replica symmetry broken (RSB) phases but also to provide additional information on the nature of the latter phase.

In order to analyse the nature of a phase transition between the RS phase and a replica-symmetry broken (RSB) phase, we will analyse the stationarity equations associated to the optimisation problem in Eq. \eqref{GS_opt_prob_2}. The stationarity condition obtained by optimising with respect to $l_0$ and $\nu$ are given in Eq. \eqref{nu_eq} and \eqref{l_0_eq} respectively. The equations obtained by optimising with respect to $f(t,h)$ and the boundary value $f(0,h)$ yield the partial differential equation Eq. \eqref{Parisi_eqA} and boundary condition in Eq. \eqref{Parisi_eq_bc}. Further, the equations obtained by optimising with respect to $P(t,h)$ and the boundary value $P(1,h)$ yield the Parisi differential equation and boundary condition in Eq. \eqref{Parisi_eqB}. The remaining stationarity equation obtained by optimising with respect to $w(\tau)$ for $\tau\in[0,l]$ takes the form
\begin{align}
    &0=\frac{l_0}{\lambda_0(0)^2}\left(\frac{2\mu}{\lambda_0(0)}-1\right)+2\mu \int_{\tau}^l \frac{dt}{\lambda_0(t)^3}-(1-\alpha)\int_{\tau}^l \frac{dt}{\lambda_0(t)^2}-\alpha \Ep{\int P(\tau,h)(\partial_{h} f(\tau,h))^2\,dh}\nn\\
    &=\frac{l_0}{\mu}+2\mu \int_{\tau}^l \frac{dt}{\lambda_0(t)^3}-(1-\alpha)\int_{\tau}^l \frac{dt}{\lambda_0(t)^2}-\alpha \Ep{\int P(\tau,h)(\partial_{h} f(\tau,h))^2\,dh}\label{sp_w}\\
    &=(1-\alpha)\int_{0}^{\tau} \frac{dt}{\lambda_0(t)^2}-2\mu \int_{0}^{\tau} \frac{dt}{\lambda_0(t)^3}+\alpha \Ep{\int \left[P(\tau,h)(\partial_{h} f(\tau,h))^2-P(\tau,l)(\partial_{h} \epsilon_{\min}(\nu,h))^2\right]\,dh}\;\nn
\end{align}
and that by optimising with respect to $l$ reads
\begin{align}
    &0=w(l)\left(\frac{l_0 }{\lambda_0(0)^2}\left(\frac{2\mu}{\lambda_0(0)}-1\right)+2\mu \int_{0}^l \frac{dt}{\lambda_0(t)^3}-(1-\alpha)\int_{0}^l \frac{dt}{\lambda_0(t)^2}-\alpha \Ep{\int P(l,h)(\partial_{h}\epsilon_{\min}(\nu,h))^2\,dh}\right)\nn\\
    &+\frac{\mu}{\nu^2}-\frac{1-\alpha}{\nu}+\alpha \Ep{\int P(l,h)\partial_{h}^2\epsilon_{\min}(\nu,h)\,dh}
    =\frac{\mu}{\nu^2}-\frac{1-\alpha}{\nu}+\alpha \Ep{\int P(l,h)\partial_{h}^2\epsilon_{\min}(\nu,h)\,dh}\;,\label{sp_l}
\end{align}
where we have used the identities 
\be
\partial_{\nu}\epsilon_{\min}(\nu,h)=\frac{1}{2}z_{\min}(\nu,h)^2\;,\;\;\partial_{h}\epsilon_{\min}(\nu,h)=-z_{\min}(\nu,h)\;.
\ee
Note that in the RS phase, Eq. \eqref{sp_l} takes the particularly simple form
\be
\Ep{\frac{\mu}{\mu+\phi''[z_{\min}(\mu)]}}=1\;,
\ee
which indeed holds throughout this phase (see Appendix \ref{replicon}).

On approaching the transition between a RS and a RSB phase, a RSB solution with a positive Parisi function $w(t)>0$ holds in an arbitrary small but finite interval $t\in[0,l]$ with $l>0$, and approaches the RS solution, with $l=0$ and $w(t)=0$. For the RSB solution the stationarity equation Eq. \eqref{sp_w} should hold throughout the interval $\tau\in [0,l]$ and one can compute properties of the solution in the vicinity of the phase transition by taking derivatives with respect to the parameter $\tau$ and inserting the RS expressions for the different parameters \cite{parisi_urbani_zamponi2020book}. In particular, taking the first derivative with respect to $\tau$, one obtains
\begin{align}
 &-\alpha  \partial_{\tau}\Ep{\int P(\tau,h)(\partial_{h} f(\tau,h))^2\,dh}-\frac{2\mu}{\lambda_0(\tau)^3}+\frac{1-\alpha}{\lambda_0(\tau)^2}\nn\\
 &=\alpha \Ep{\int P(\tau,h)(\partial_{h}^2 f(\tau,h))^2\,dh}-\frac{2\mu}{\lambda_0(\tau)^3}+\frac{1-\alpha}{\lambda_0(\tau)^2}=0\;,\label{trans_eq}    
\end{align}
where we have used the identities derived in Appendix \ref{id_FRSB} to simplify this expression. Inserting here the RS solution allows to reproduce the criterion for a vanishing replicon eigenvalue, derived in Appendix \ref{replicon}:
\be \label{replicon2}
\lambda_{\rm RS}=\alpha\mu^2\Ep{\left(\partial_{h}^2 \epsilon_{\min}(\nu,h)\right)_{h=0}^2}-(1+\alpha)=\alpha\Ep{\left(\frac{\mu}{\mu+\phi''[z_{\min}(\mu)]}\right)^2}-(1+\alpha)=0\;.
\ee
Solving the equation above allows to determine the location of the de Almeida-Thouless line corresponding to a continuous transition between a RS and RSB phase. Note however that for those models in this class  where $p_0(t)$ has unbounded support 
(in particular, for the Gaussian density) the expectation in Eq. \eqref{replicon2} diverges, so that the replicon eigenvalue is always positive, hence  the optimal solution is always in a RSB phase. Let us now consider the case where a transition between the two phases indeed occurs and analyse some properties of the transition.

In order to obtain additional information on the nature of the transition, it is expedient to compute successive derivative of Eq. \eqref{trans_eq}. Taking the first derivative with respect to $\tau$, one obtains

\begin{align}
  2w(\tau)\left(\frac{3\mu}{\lambda_0(\tau)^4}-\frac{1-\alpha}{\lambda_0(\tau)^3}-\alpha\Ep{\int P(\tau,h)\left(\partial_{h}^2 f(\tau,h)\right)^3\,dh}\right)
  &=-\alpha\Ep{\int P(\tau,h)\left(\partial_{h}^3 f(\tau,h)\right)^2\,dh}\;, \label{breaking_point} 
\end{align}
and upon inserting here the RS expression one gets the so-called (rescaled) breaking point\cite{parisi_urbani_zamponi2020book} , which can be determined as
\be
w_{\rm AT}=-\frac{\alpha\mu^3\Ep{\left(\partial_{h}^3 \epsilon_{\min}(\nu,h)\right)_{h=0}^2}}{2\left(\alpha\mu^3\Ep{\left(\partial_{h}^2 \epsilon_{\min}(\nu,h)\right)_{h=0}^3}+\alpha+2\right)}=\frac{\alpha\mu^3\Ep{{\cal C}_3(\mu)^2}}{2\left(\alpha\mu^3\Ep{{\cal C}_2(\mu)^3}-(\alpha+2)\right)}\;.\label{breaking_point2}
\ee
Here and in the following, the functions ${\cal C}_k(\mu)=-\partial_h^k \epsilon(\mu,0)$ are defined in Eq. \eqref{C_k_cum} and correspond to rescaled cumulants of $z_{\min}$. The continuous transition to RSB phase is only possible when the breaking point is positive: $w_{\rm AT}>0$. In the opposite case $w_{\rm AT}<0$ a discontinuous transition to a RSB phase is known to develop before the continuous transition \cite{parisi_urbani_zamponi2020book}. This type of transition will be considered in the next section. Finally, taking an additional derivative in Eq. \eqref{breaking_point}, one obtains
\begin{align}
    &2w'(\tau)\left(\frac{3\mu}{\lambda_0(\tau)^4}-\frac{1-\alpha}{\lambda_0(\tau)^3}-\alpha\Ep{\int P(\tau,h)\left(\partial_h^2 f(\tau,h)\right)^3\,dh}\right)\\
  =&- 6w(\tau)^2 \alpha\Ep{\int P(\tau,h)\left(\partial_h^2 f(\tau,h)\right)^4\,dh}+ 12w(\tau)\alpha\Ep{\int P(\tau,h)\partial_h^2 f(\tau,h)\left(\partial_h^3 f(\tau,h)\right)^2\,dh}\nn\\
  &-\alpha\Ep{\int P(\tau,h)\left(\partial_h^4 f(\tau,h)\right)^2\,dh}+6w(\tau)\left(\frac{4\mu}{\lambda_0(\tau)^5}-\frac{1-\alpha}{\lambda_0(\tau)^4}\right)\;.\nn
\end{align}
Inserting the RS ansatz one then finds the first derivative of $w(\tau)$ at the transition to be equal to 
\be
w_{\rm AT}'=\frac{\alpha\mu^4\left(\mathbb{E}\left[{\cal C}_4(\mu)^2\right]-12 w_{\rm AT}\,\mathbb{E}\left[{\cal C}_3(\mu)^2{\cal C}_2(\mu)\right]+6w_{\rm AT}^2\,\mathbb{E}\left[{\cal C}_2(\mu)^4\right]\right)-6(\alpha+3)w_{\rm AT}^2}{2\left(\alpha\mu^3\,\mathbb{E}\left[{\cal C}_2(\mu)^3\right]-(\alpha+2)\right)}\;.\label{slope}
\ee
The replica theory requires the function $w(\tau)$ to be non-decreasing. A positive value $w_{\rm AT}'>0$ of this quantity indicates that the solution $w(\tau)$ is actually increasing in the vicinity of the transition. It thus indicates that the corresponding transition should be from a RS to the so-called  full replica symmetry breaking (FRSB) phase. If it happens that $w_{\rm AT}'<0$ this fact reveals inconsistency in the derivation indicating that the solution below the transition is actually one-step RSB (1RSB). In that case the function $w(\tau)$ is piece-wise constant and one cannot obtain more information by taking derivatives with respect to $\tau$ \cite{parisi_urbani_zamponi2020book}.

This computation concludes our analysis of the continuous RSB transition. We will now consider in more detail the properties of the 1RSB phase and the occurrence of discontinuous transitions from a RS to a 1RSB phase.

\subsection{One-step replica symmetry breaking phase and discontinuous transition}

 Recalling that the functions $f(t,h)$ and $P(t,h)$ satisfy the differential equations \eqref{Parisi_eqA} and \eqref{Parisi_eqB} correspondingly, one can define respectively the functions
\be
G(t,h)=e^{w(t)f(t,h)}\;,\;\;K(t,h)=P(t,h)e^{-w(t)f(t,h)}\;,\;\;t\in[0,l]\;,\;\;h\in \mathbb{R}\;,
\ee
which in turn satisfy the following partial differential equations:
\begin{align}
\partial_t G(t,h)&=-\frac{1}{2}\partial_{h}^2 G(t,h)+\frac{w'(t)}{w(t)}G(t,h)\ln G(t,h)\;,\\
 \partial_t K(t,h)&=\frac{1}{2}\partial_{h}^2 K(t,h)-w'(t)f(t,h)\,K(t,h)\;.   
\end{align}
The 1RSB phase is characterized by the function $w(t)$ retaining a constant value $w(t)=m>0$ for all $t\in [0,l]$.  The functions $G(t,h)$ and $K(t,h)$ then simply satisfy a backward and forward diffusion equation, respectively. Using the boundary condition at $t=l$ one finds
\be
G(t,h)=\int e^{-\frac{(h-h_0)^2}{2(l-t)}-m \epsilon_{\min}(\nu,h_0)}\,\frac{dh_0}{\sqrt{2\pi(l-t)}}\;,\;\;t\in(0,l)\;,\;\;h\in \mathbb{R}\;,
\ee
hence, the function
\be
f(0_+,h)=\frac{1}{m}\ln \int e^{-\frac{(h_0-h)^2}{2l}-m \epsilon_{\min}(\nu,h_0)}\,\frac{dh_0}{\sqrt{2\pi l}}\;.
\ee
On the other hand, we have
\be
K(t,h)=\int e^{-\frac{(h-h_0)^2}{2t}-m f(0_+,h_0)}\delta(h_0)\, \frac{dh_0}{\sqrt{2\pi t}}=\sqrt{\frac{l}{t}}\frac{\displaystyle e^{-\frac{h^2}{2t}}}{\int dh_1\displaystyle e^{-\frac{h_1^2}{2l}-m\epsilon_{\min}(\nu,h_1)}}\;,\;\;t\in(0,l)\;,\;\;h\in \mathbb{R}\;.
\ee
Then the function $P(t,h)$ can be obtained as
\be
P(t,h)=\sqrt{\frac{l}{2\pi t(l-t)}}e^{-\frac{h^2}{2t}}\int \frac{\displaystyle e^{-\frac{(h-h_0)^2}{2(l-t)}-m\epsilon_{\min}(\nu,h_0)}}{\int  \displaystyle e^{-\frac{h_1^2}{2l}-m\epsilon_{\min}(\nu,h_1)}\,dh_1}\,dh_0\;,\;\;t\in(0,l)\;,\;\;h\in \mathbb{R}\;.
\ee
One can check in particular that $\int P(t,h)\,dh=1$ as expected. This probability distribution $P(t,h)$ is standardly interpreted as the distribution of the so-called {\it local fields}\cite{sommers1984distribution}.

The optimisation with respect to $l_0$ in Eq. \eqref{nu_eq} and $\nu$ in Eq. \eqref{l_0_eq} yield the simple results
\be
\nu=\mu+m\,l\;,\;\;\frac{l_0}{\mu^2}+\frac{l(\mu+\alpha(\mu+m l))}{\mu^2(\mu+m l)^2}=\alpha\Ep{\frac{\int e^{-\frac{h^2}{2l}-m\epsilon_{\min}(\mu+m l,h)}\,z_{\min}(\mu+m l,h)^2\,dh}{\int \,e^{-\frac{h_0^2}{2l}-m\epsilon_{\min}(\mu+m l,h_0)}\,dh_0}}\;.
\ee
One recovers the RS expression in the limit $l\to 0$. In the limit $m\to 0$ a similar result holds up to the modification $l_0\to l_0+l$. Taking into account these expressions the ground-state energy difference between the RS and 1RSB phase can be expressed as an optimisation problem solely depending on $l\geq 0$ and $m\geq 0$. It reads
\begin{align}
    \Delta e_{\rm 1RSB}=&\underset{l,m\geq 0}{\rm max}\Phi_{\mu}(m,l)\;,\\
    \Phi_{\mu}(m,l)=&\frac{1}{2}\left(\frac{l}{\mu+ml}-\frac{1-\alpha}{m}\ln\left(1+\frac{m l}{\mu}\right)\right)\nn\\
    &-\frac{\alpha}{m}\Ep{\ln\left(\int e^{-\frac{h^2}{2l}-m\epsilon_{\min}(\mu+m l,h)} \frac{dh}{\sqrt{2\pi l}}\right)+m\epsilon_{\min}(\mu)}\;.\label{d1RSB_1}
\end{align}
The stationarity conditions for the function to be optimised  read
\begin{align}
    &\frac{\alpha \mu-(1-\alpha)m l}{(\mu+m l)^2}\nn\\
    &+\alpha\Ep{\frac{\displaystyle \int e^{-\frac{h^2}{2l}-m\epsilon_{\min}(\mu+m l,h)}\left(m z_{\min}(\mu+m l,h)^2-\frac{h}{l}z_{\min}(\mu+m l,h)\right)\,dh }{\displaystyle \int e^{-\frac{h_0^2}{2l}-m\epsilon_{\min}(\mu+m l,h_0)}\,dh_0}}=0\;,\\
    &-\frac{l\left(m l(2-\alpha)+\mu(1-\alpha)\right)}{2m(\mu+m l)^2}+\frac{1-\alpha}{m^2}\ln\left(1+\frac{m l}{\mu}\right)+\frac{\alpha}{m^2}\Ep{\ln\left(\int e^{-\frac{h^2}{2l}-m\epsilon_{\min}(\mu+m l,h)}\right)\frac{dh}{\sqrt{2\pi l}}}\nn\\
    &+\frac{\alpha}{m}\Ep{\frac{\displaystyle \int e^{-\frac{h^2}{2l}-m\epsilon_{\min}(\mu+m l,h)}\left(\epsilon_{\min}(\mu+m l,h)+\frac{m l}{2}z_{\min}(\mu+m l,h)^2\right)\,dh }{\displaystyle \int e^{-\frac{h_0^2}{2l}-m\epsilon_{\min}(\mu+m l,h_0)}\,dh_0}}=0\;.
\end{align}

A phase transition is expected to occur in the model whenever the ground-state energy difference vanishes, which occurs either via $l\to 0$ or via $m\to 0$. Expanding the expression of $\Phi_{\mu}(m,l)$ up to second order in $l$, one obtains
\be
\Phi_{\mu}(m,l)=\frac{m l^2}{4\mu^2}\left(\alpha\Ep{\frac{\mu^2}{\left(\mu+\phi''[z_{\min}(\mu)]\right)^{2}}}-(1+\alpha)\right)+O(l^3)=\lambda_{\rm RS}\frac{m l^2}{4\mu^2}+O(l^3)\;,
\ee
where $\lambda_{\rm RS}$ is the replicon eigenvalue. Thus, the criterion for a stable 1RSB phase near the transition (i.e. when $l\to 0$) corresponds to $\Delta e_{\rm 1RSB}>0$ and coincides with the criterion for an unstable RS phase. Similarly, expanding in $m$ yields
\begin{align}
 \Phi_{\mu}(m,l)=&m A(l)+O(m^2)\;,\\
A(l)=&\alpha\,\Ep{\left(\int e^{-\frac{h^2}{2l}}\epsilon_{\min}(\mu,h)\frac{dh}{\sqrt{2\pi l}}\right)^2-\int e^{-\frac{h^2}{2l}}\left(\epsilon_{\min}^2(\mu,h)-l\,z_{\min}^2(\mu,h)\right)\frac{dh}{\sqrt{2\pi l}}}\\
&-\frac{l^2(1+\alpha)}{4\mu^2}\;.   \nn
\end{align}
The criterion for this discontinuous transition is more stringent than for the continuous transition as can be checked by expanding the function  $A(l)\approx \lambda_{\rm RS}l^2/(4\mu^2)$ in the limit $l\to 0$.

Note finally that in the 1RSB phase the value for overlaps in the ground state can be recovered from the optimal values of $l_0,l,m$ as
\begin{align}
 r_0&=\frac{l_0}{\mu^2}\;,\;\;r_1-r_0=\frac{l}{\mu(\mu+m l)}\;,\;\;v=\lim_{\beta\to \infty}\beta(r_d-r_1)=\frac{1}{\mu+m l}\;.
\end{align}

To conclude our analysis of the phase diagram, we provide in the next subsection the criterion for the transition between distinct RSB phases.

\subsection{Gardner transition}

Let us finally consider a possible continuous Gardner transition from a 1RSB phase to a FRSB phase. At such a transition both solutions should approach each other, similarly to the previously studied case of the continuous transition from RS to the 1RSB phase. The transition then again manifests itself by the emergence of a marginal eigenvalue of the quadratic form controlling the stability of the 1RSB expression for the ground-state energy. The location of this transition can be obtained by inserting the 1RSB solution in Eq. \eqref{trans_eq}. Taking in particular $\tau\to 0_+$, one obtains the expression for the first eigenvalue to be given by
\begin{align}
\frac{\lambda_1}{\mu^2}=&\alpha\Ep{\left(\int P(l,h)\,\partial_{h}^2\epsilon_{\min}\,dh-m\left[\int P(l,h)\,(\partial_{h}\epsilon_{\min})^2\,dh
-\left(\int P(l,h)\,\partial_{h}\epsilon_{\min}\,dh\right)^2\right]\right)^2}\nn\\
&-\frac{1+\alpha}{\mu^2}\;,\label{lamb_1}
\end{align}
where $\epsilon_{\min}\equiv \epsilon_{\min}(\mu+m l,h)$. On the other hand, taking $\tau \to l_-$ yields instead
\be
\frac{\lambda_2}{\mu^2}=\alpha\Ep{\int P(l,h)\,\left(\partial_{h}^2\epsilon_{\min}\right)^2\,dh}-\frac{2}{(\mu+m l)^3}+\frac{1-\alpha}{(\mu+m l)^2}\;,\label{lamb_2}
\ee
where $P(l,h)$ is given by
\be
P(l,h)=\frac{\displaystyle e^{-\frac{h^2}{2l}-m \epsilon_{\min}(\mu+m l,h)}}{\displaystyle
\int e^{-\frac{h_0^2}{2l}-m \epsilon_{\min}(\mu+m l,h_0)}\,dh_0}\;.
\ee
One can easily check that $\lambda_{1}=\lambda_{2}=\lambda_{\rm RS}$ when one takes either $m\to 0$ or $l\to 0$.

\section{Conclusion} \label{sec_conclu}

In this article, we have introduced and analysed some statistical properties of a class of models of high-dimensional random landscapes of the form ${\cal H}({\bf x})=\frac{\mu}{2}{\bf x}^2+\sum_{l=1}^M \phi_l({\bf k}_l\cdot {\bf x}), \, \, \mu>0$ defined in a Euclidean space ${\bf x}\in \mathbb{R}^N$, with both the functions $\phi_l(z)$ and vectors ${\bf k}_l$ chosen to be random. An important example of such landscape
describes superposition of $M$ plane waves with random amplitudes, directions of the wavevectors, and phases, further confined by a parabolic potential of curvature $\mu$. 
 Our main efforts have been directed towards analysing  the landscape features in the limit $N\to \infty, M\to \infty$ keeping $\alpha=M/N$ finite. In such a limit we find (i) 
the expression for the mean (and simultaneously typical) value of the deepest landscape minimum (the ground-state energy), expressed via the solution to a Parisi-type optimisation problem and (ii) the rates of asymptotic exponential growth with $N$ of the mean number of all critical points and of local minima known as the annealed complexities.  Note that although the complexity of minima is well-defined also for random landscapes defined on discrete phase-space, such as the celebrated Sherrington-Kirkpatrick (SK) model,  the methods to compute it seem not yet available. The class of models considered here is quite versatile, displaying rich qualitatively different behaviour depending on the probability density function $p_0(t)$ of the curvature  $t=\phi''(z)$ of the random potential $\phi(z)$ featuring in Eq. \eqref{H}:
\begin{itemize}
    \item If  $p_0(t)$ has bounded support, the ground-state energy changes its properties with varying curvature confinement parameter $\mu$ and displays a transition from the replica-symmetric to a replica-symmetry broken phase. This replica-symmetry breaking transition coincides with the topology trivialisation transition, separating a trivial phase with zero annealed complexities to a complex phase with positive annealed complexities. Such properties are qualitatively similar to what is observed in the zero-dimensional toy model for elastic manifold defined in Eqs. (\ref{H}) and (\ref{GRF}) and studied notably in \cite{fyodorov2004complexity,fyodorov2007classical,fyodorov2007replica,BenArousBourgadeMcKenna23}.
    \item If the support of $p_0(t)$ is unbounded, the ground-state energy always corresponds to the replica-symmetry broken phase for any $\mu$, not dissimilar to the case of the celebrated Sherrington-Kirkpatrick model in a magnetic field. The random landscape is always complex, with positive annealed complexities for all parameters. 
\end{itemize}
Our analysis  can be certainly extended in several directions which we now briefly mention. The most immediate would be to compute the annealed complexities at fixed energy and to apply the replicated Kac-Rice method to extract the quenched complexities (see \cite{ros2019complex,ros2023generalized,ros2023quenched}). Another natural direction would be to explore the fluctuations of the ground-state energy by computing its large deviation function (see \cite{lacroix2024replica} for the case of spherical model). Completely unexplored are dynamical properties  for models in this class, with expected 
rich glassy phenomenology, in particular, the ageing phenomena, see \cite{crisanti2023dynamical}.

Finally, we hope that the optimisation functional for the mean ground state derived by us using the replica method could be amenable to derivation by rigorous methods of modern probability theory.  

\acknowledgements 
We would like to thank Pierre Le Doussal for his interest in the work and participation at the early stages of  the
research reported in this paper, and Igor Wigman for useful remarks on the draft of the paper. The research by Bertrand Lacroix-A-Chez-Toine and Yan V. Fyodorov was supported by the EPSRC under Grant No. EP/V002473/1 (Random
Hessians and Jacobians: theory and applications).

\appendix

\section{Simplification of the smallest inverse overlap using the stationarity}\label{app_q0}

In this appendix, we provide details on the method to simplify the term involving $q_0$ using stationarity of the disordered potential $\phi(z)$, which is presented in Eq. \eqref{removing_q0}. First, one can write for $k$ replica symmetry breaking
\begin{align}
 &\mathbb{E}_{\phi}\left[\int \frac{dh_0}{\sqrt{-2\pi  q_0}}e^{\frac{h_0^2}{2 q_0}+n g_0(h_0,\nu)}\right]=\int \frac{dh_0}{\sqrt{-2\pi q_0}}e^{\frac{h_0^2}{2 q_0}}\mathbb{E}_{\phi}\left[e^{n g_0(h_0,\nu)}\right]\\ 
 &g_{j-1}(h_{j-1},\nu)=\frac{1}{m_{j-1}}\ln \int \frac{dh_{j}}{\sqrt{2\pi(p_j-p_{j-1})}}e^{-\frac{(h_{j-1}-h_j)^2}{2(p_j-p_{j-1})}+m_{j-1} g_j(h_j,\nu)}\\
 &g_{k+1}(h_k,\nu)=\frac{1}{m_{k+1}}\ln\int dz\,e^{-\beta\left[\frac{\nu}{2}z^2+\phi(z)\right]+h_k z}\;.
\end{align}
Rewriting all terms in a single exponential, its argument reads up to a global minus sign
\begin{align}
    \sum_{j=2}^{k-1}\frac{(h_{j-1}-h_j)^2}{2m_j(p_j-p_{j-1})}+\frac{h_1^2}{2m_1 p_1}-\frac{h_0^2}{2n q_0}+\frac{\beta\nu}{2}z^2+\beta\phi(z)-(h_k+h_0)z\;.
\end{align}
Let us then apply the following successive change of variables in the integrals $h_j\to h_j+h_0\sum_{i=1}^j s_j$, where the values of each $s_j$ will be set in the next steps. It yields
\begin{align}
    &\sum_{j=2}^{k-1}\left[\frac{(h_{j-1}-h_j)^2}{2m_j(p_j-p_{j-1})}+\frac{s_j(h_{j-1}-h_j)}{m_j(p_j-p_{j-1})}+\frac{s_j^2}{2m_j(p_j-p_{j-1})}\right]\\
    &-\frac{h_0^2}{2n q_0}+\frac{h_1^2+2s_1 h_0 h_1 +s_1^2 h_0^2}{2m_1 p_1}+\frac{\beta\nu}{2}z^2+\beta\phi(z)-(h_k+h_0)z\nn\\
=&\sum_{j=2}^{k-1}\left[\frac{(h_{j-1}-h_j)^2}{2m_j(p_j-p_{j-1})}+\frac{s_j^2\,h_0^2}{2m_j(p_j-p_{j-1})}\right]+\frac{s_1^2 h_0^2}{2m_1 p_1}+h_0\sum_{j=2}^{k-1}h_j\left[\frac{s_{j+1}}{m_{j+1}(p_{j+1}-p_{j})}-\frac{s_j}{m_j(p_j-p_{j-1})}\right]\nn\\
&-\frac{h_0^2}{2n q_0}+\frac{h_1^2}{2m_1 p_1}+h_0 h_1\left[\frac{s_2}{m_2 (p_2-p_1)}-\frac{s_1}{m_1 p_1}\right]+\frac{\beta\nu}{2}z^2+\beta\phi(z)-\left[h_k+h_0\left(1+\sum_{j=1}^k s_j\right)\right]z
\end{align}
Now, we apply an additional change of variable $z\to z+h_0/(\beta\nu)\left(1+\sum_{j=1}^k s_j\right)$ and use the stationarity of the disordered potential $\phi(z)$, yielding
\begin{align}
&\sum_{j=2}^{k-1}\frac{(h_{j-1}-h_j)^2}{2m_j(p_j-p_{j-1})}+\frac{h_1^2}{2m_1 p_1}+\frac{\beta\nu}{2}z^2+\beta\phi(z)-h_k z+\left[\frac{s_k}{m_k(p_k-p_{k-1})}-\frac{1}{\beta\nu}\left(1+\sum_{j=1}^k s_j\right)\right]h_0 h_k\nn\\
&+h_0\sum_{j=2}^{k-1}h_j\left[\frac{s_{j+1}}{m_{j+1}(p_{j+1}-p_{j})}-\frac{s_j}{m_j(p_j-p_{j-1})}\right]+h_0 h_1\left[\frac{s_2}{m_2 (p_2-p_1)}-\frac{s_1}{m_1 p_1}\right]\label{exponential_coeff}\\
&+\frac{h_0^2}{2}\left[-\frac{1}{n q_0}-\frac{2}{\beta\nu}\left(1+\sum_{j=1}^k s_j\right)^2+\sum_{j=1}^k \frac{s_j^2}{m_j(p_j-p_{j-1})}\right]\;.\nn
\end{align}
The values of the different $s_j$'s is now set by cancelling the linear terms in $h_j$'s for $j=1,\cdots,k-1$
\be
s_{j}=\frac{m_j(p_j-p_{j-1})}{m_{j+1}(p_{j+1}-p_{j})}s_{j+1}=\frac{m_j(p_j-p_{j-1})}{m_{k}(p_{k}-p_{k-1})}s_k\;,\;\;s_1=\frac{m_1 p_1}{m_{k}(p_{k}-p_{k-1})}s_k\;.
\ee
Finally, cancelling the linear term in $h_k$, one obtains
\begin{align}
&\frac{s_k}{m_k(p_k-p_{k-1})}-\frac{1}{\beta\nu}\left(1+s_k\sum_{j=1}^k \frac{m_j(p_j-p_{j-1})}{m_k(p_k-p_{k-1})}\right)=0\;,\;\;s_k=\frac{m_k(p_k-p_{k-1})}{\lambda_1}\;,\\
&s_j=\frac{m_j(p_j-p_{j-1})}{\lambda_1}\;,\\
&\lambda_1=\beta\nu-\sum_{j=1}^k m_j(p_j-p_{j-1})\;,\;\;\frac{1}{\beta\nu}\left(1+\sum_{j=1}^k s_j\right)=\frac{1}{\lambda_1}\;.
\end{align}
Inserting these identities in Eq. \eqref{exponential_coeff}, one obtains
\begin{align}
&\sum_{j=2}^{k-1}\frac{(h_{j-1}-h_j)^2}{2m_j(p_j-p_{j-1})}+\frac{h_1^2}{2m_1 p_1}+\frac{\beta\nu}{2}z^2+\beta\phi(z)-h_k z-\frac{h_0^2}{2}\left(\frac{1}{n q_0}+\frac{1}{\lambda_1}\right)\;.
\end{align}
In particular the field $h_0$ is no longer coupled to any other integration variable and one can now compute explicitly the associated Gaussian integral
\be
\int \frac{dh_0}{\sqrt{-2\pi q_0}}e^{\frac{h_0^2}{2}\left(\frac{1}{\bar q_0}+\frac{n}{\lambda_1}\right)}=\left(1+\frac{n q_0}{\lambda_1}\right)^{-\frac{1}{2}}\;.
\ee
Finally, one obtains the result displayed in the second line of Eq. \eqref{removing_q0}.

\section{Stability of the RS solution} \label{replicon}

In this section, we derive the expression of the replicon eigenvalue in Eq. \eqref{replicon_eigv}, delimiting the continuous transition from the RS to a RSB phase, following a similar procedure as in the original article by de Almeida and Thouless \cite{AT78}.
To this end, we expand the function $\Sigma_{n}(Q+\epsilon \,\eta)$ appearing in Eq. \eqref{action} up to second order in powers of $\epsilon$. It reads
\begin{align}
\Sigma_{n}(Q+&\epsilon\,\eta)=\Sigma_{n}(Q)+\epsilon\sum_{a,b=1}^n W_{ab}\,\eta_{ab}+\frac{\epsilon^2}{2}\sum_{a,b,c,d=1}^n G_{abcd}\,\eta_{ab}\eta_{cd}+o(\epsilon^2)\;,\\
    W_{ab}=&\partial_{Q_{ab}}\Sigma_{n}(Q)=\frac{\beta\mu}{2}(Q^{-2})_{ab}-\frac{1-\alpha}{2}(Q^{-1})_{ab}-\frac{\alpha}{2}\frac{\mathbb{E}_{\phi}\left[\int d{\bf z}\,z_a\,z_b\,e^{-\frac{1}{2}{\bf z}Q{\bf z}-\beta\sum_{m=1}^n \phi(z_m)}\right]}{\mathbb{E}_{\phi}\left[\int d{\bf z}\,e^{-\frac{1}{2}{\bf z}Q{\bf z}-\beta\sum_{m=1}^n \phi(z_m)}\right]}\;,\\
    G_{abcd}=&\partial_{Q_{ab},Q_{cd}}^2\Sigma_{n}(Q)=-\frac{\beta\mu}{2}\,\left[(Q^{-1})_{bc}(Q^{-2})_{da}+(Q^{-2})_{bc}(Q^{-1})_{da}\right]\\
    &+\frac{1-\alpha}{2}(Q^{-1})_{bc}(Q^{-1})_{da}+\frac{\alpha}{4}\frac{\mathbb{E}_{\phi}\left[\int d{\bf z}\,z_a\,z_b\,z_c\,z_d\,e^{-\frac{1}{2}{\bf z}Q{\bf z}-\beta\sum_{m=1}^n \phi(z_m)}\right]}{\mathbb{E}_{\phi}\left[\int d{\bf z}\,e^{-\frac{1}{2}{\bf z}Q{\bf z}-\beta\sum_{m=1}^n \phi(z_m)}\right]}\nn\\
    &-\frac{\alpha}{4}\left(\frac{\mathbb{E}_{\phi}\left[\int d{\bf z}\,z_a\,z_b\,e^{-\frac{1}{2}{\bf z}Q{\bf z}-\beta\sum_{m=1}^n \phi(z_m)}\right]}{\mathbb{E}_{\phi}\left[\int d{\bf z}\,e^{-\frac{1}{2}{\bf z}Q{\bf z}-\beta\sum_{m=1}^n \phi(z_m)}\right]}\right)\left(\frac{\mathbb{E}_{\phi}\left[\int d{\bf z}\,z_c\,z_d\,e^{-\frac{1}{2}{\bf z}Q{\bf z}-\beta\sum_{m=1}^n \phi(z_m)}\right]}{\mathbb{E}_{\phi}\left[\int d{\bf z}\,e^{-\frac{1}{2}{\bf z}Q{\bf z}-\beta\sum_{m=1}^n \phi(z_m)}\right]}\right)\;.\nn
\end{align}
To check the stability of the RS ansatz, one should compute the eigenvalues of the second-order term, i.e. finding the solutions of
\begin{align}
\Lambda_n\, \eta_{aa}=&\sum_{c,d}G_{aacd}\,\eta_{cd}=\sum_{c}G_{aacc}\,\eta_{cc}+\sum_{c\neq d}G_{aacd}\,\eta_{cd}\label{eq_eig_aa}\\
=&G_{aaaa}\,\eta_{aa}+\sum_{c\neq a}G_{aacc}\,\eta_{cc}+\sum_{c\neq a}G_{aaca}\,\eta_{ca}+\sum_{d\neq a}G_{aaad}\,\eta_{ad}+\sum_{c\neq d,c\neq a,d\neq a}G_{aacd}\,\eta_{cd}\;,\nn\\
=&\left(G_{aaaa}-G_{aabb}\right)\eta_{aa}+G_{aabb}\sum_{b}\eta_{bb}+\left(G_{aaba}+G_{aaab}-2G_{aacd}\right)\sum_{b}\eta_{ab}(1-\delta_{ab})+G_{aacd}\sum_{c\neq d}\eta_{cd}\;,\nn\\
 \Lambda_n\, \eta_{ab}=&\sum_{c,d}G_{abcd}\,\eta_{cd}=\sum_{c}G_{abcc}\,\eta_{cc}+\sum_{c\neq d}G_{abcd}\,\eta_{cd}\label{eq_eig_ab}\\
 =&G_{abaa}\,\eta_{aa}+G_{abbb}\,\eta_{bb}+\sum_{c\neq a,b}G_{abcc}\,\eta_{cc}+G_{abab}\,\eta_{ab}+G_{abba}\,\eta_{ba}\nn\\
 &+\sum_{d\neq a,b}G_{abad}\,\eta_{ad}+\sum_{d\neq a,b}G_{abbd}\,\eta_{bd}+\sum_{c\neq a,b}G_{abca}\,\eta_{ca}+\sum_{c\neq a,b}G_{abcb}\,\eta_{cb}+\sum_{c\neq d,c\neq a,b,d\neq a,b}G_{abcd}\,\eta_{cd}\nn\\
=&\left(G_{aaab}-G_{aabc}\right)\,\eta_{aa}+\left(G_{aaba}-G_{aabc}\right)\,\eta_{bb}+G_{aabc}\sum_{c}\eta_{cc}\nn\\
&+\left(G_{abba}+G_{abab}+2G_{abcd}-G_{abac}-G_{abbc}-G_{abca}-G_{abcb}\right)\eta_{ab}\nn\\
&+\left(G_{abac}+G_{abca}-2G_{abcd}\right)\sum_{c}\eta_{ac}(1-\delta_{ac})+\left(G_{abbc}+G_{abcb}-2G_{abcd}\right)\sum_{c}\eta_{bc}(1-\delta_{bc})\nn\\
&+G_{abcd}\sum_{c\neq d}\eta_{cd}\;,\;\;a\neq b\;.\nn
\end{align}
One can compute explicitly all the different terms in the RS phase, using the identities
\begin{align}
   &(Q^{-1})_{ab}=r_0+(r_d-r_0)\delta_{ab}\;,\;\;(Q^{-2})_{ab}=R_0+(R_d-R_0)\delta_{ab}\\
   &\frac{\mathbb{E}_{\phi}\left[\int d{\bf z}\,z_a\,z_b\,z_c\,z_d\,e^{-\frac{1}{2}{\bf z}Q{\bf z}-\beta\sum_{m=1}^n \phi(z_m)}\right]}{\mathbb{E}_{\phi}\left[\int d{\bf z}\,e^{-\frac{1}{2}{\bf z}Q{\bf z}-\beta\sum_{m=1}^n \phi(z_m)}\right]}=\frac{\mathbb{E}_{\phi}\left[\int d{\bf z}\,z_a\,z_b\,z_c\,z_d\,e^{-\beta\sum_{m=1}^n\left[\frac{\nu}{2}z_m^2+\phi(z_m)\right]}\right]}{\mathbb{E}_{\phi}\left[\int d{\bf z}\,e^{-\beta\sum_{m=1}^n\left[\frac{\nu}{2}z_m^2+\phi(z_m)\right]}\right]}+3r_0^2\nn\\
   &+r_0\frac{\mathbb{E}_{\phi}\left[\int d{\bf z}\,(z_a\,z_b+z_a\,z_c+z_a\,z_d+z_b\,z_c+z_b\,z_d+z_c\,z_d)\,e^{-\beta\sum_{m=1}^n\left[\frac{\nu}{2}z_m^2+\phi(z_m)\right]}\right]}{\mathbb{E}_{\phi}\left[\int d{\bf z}\,e^{-\beta\sum_{m=1}^n\left[\frac{\nu}{2}z_m^2+\phi(z_m)\right]}\right]}\;,\\
   &\frac{\mathbb{E}_{\phi}\left[\int d{\bf z}\,z_a\,z_b\,e^{-\frac{1}{2}{\bf z}Q{\bf z}-\beta\sum_{m=1}^n \phi(z_m)}\right]}{\mathbb{E}_{\phi}\left[\int d{\bf z}\,e^{-\frac{1}{2}{\bf z}Q{\bf z}-\beta\sum_{m=1}^n \phi(z_m)}\right]}=\frac{\mathbb{E}_{\phi}\left[\int d{\bf z}\,z_a\,z_b\,e^{-\beta\sum_{m=1}^n\left[\frac{\nu}{2}z_m^2+\phi(z_m)\right]}\right]}{\mathbb{E}_{\phi}\left[\int d{\bf z}\,e^{-\beta\sum_{m=1}^n\left[\frac{\nu}{2}z_m^2+\phi(z_m)\right]}\right]}+r_0\;,\\
   &r_0=-\frac{q_0}{\beta\nu(\beta\nu+n q_0)}\;.
\end{align}
In order to simplify the notations, we introduce the following partition function and thermal moments of the position of the one-dimensional disordered Hamiltonian 
\be
Z_{\nu}(\beta)=\int e^{-\beta\left(\frac{\nu}{2}z^2+\phi(z)\right)}\,dz, 
\quad \left\langle z^n\right\rangle=\frac{\int z^n\, e^{-\beta\left(\frac{\nu}{2}z^2+\phi(z)\right)}\,dz}{Z_{\nu}(\beta)}\;.  
\ee
We use below the simplified notation $Z_{\nu}\equiv Z_{\nu}(\beta)$. It is then possible to express explicitly
\begin{align}
  &G_{aaaa}=-\beta\mu R_d r_d+\frac{(1-\alpha)r_d^2+\alpha r_0^2}{2}+\frac{\alpha}{4}G_1\;,\\
  &G_{aaab}=G_{aaba}=-\frac{\beta\mu}{2}(R_d r_0+R_0 r_d)+\frac{(1-\alpha)r_d r_0+\alpha r_0^2}{2}+\frac{\alpha}{4}G_2\;,\\
  &G_{aabb}=-\beta\mu R_0 r_0+\frac{r_0^2}{2}+\frac{\alpha}{4}G_3\\
  &G_{abab}=-\beta\mu R_0 r_0+\frac{r_0^2}{2}+\frac{\alpha}{4}G_4\\
  &G_{abba}=-\beta\mu R_d r_d+\frac{(1-\alpha)r_d^2+\alpha r_0^2}{2}+\frac{\alpha}{4}G_4\\
  &G_{abac}=G_{abcb}=-\beta\mu R_0 r_0+\frac{r_0^2}{2}+\frac{\alpha}{4}G_5\\
  &G_{abca}=G_{abbc}=-\frac{\beta\mu}{2}(R_d r_0+R_0 r_d)+\frac{(1-\alpha)r_d r_0+\alpha r_0^2}{2}+\frac{\alpha}{4}G_5\\
  &G_{aabc}=-\beta\mu R_0 r_0+\frac{r_0^2}{2}+\frac{\alpha}{4}G_6\\
  &G_{abcd}=-\beta\mu R_0 r_0+\frac{r_0^2}{2}+\frac{\alpha}{4}G_7
\end{align}
where
\begin{align}
G_1&=\frac{\mathbb{E}_{\phi}\left[\moy{z^4}_{\nu}Z_{\nu}^n\right]}{\mathbb{E}_{\phi}\left[Z_{\nu}^n\right]}+4r_0\frac{\left[\moy{z^2}_{\nu}Z_{\nu}^n\right]}{\mathbb{E}_{\phi}\left[Z_{\nu}^n\right]}-\left(\frac{\mathbb{E}_{\phi}\left[\moy{z^2}_{\nu}Z_{\nu}^n\right]}{\mathbb{E}_{\phi}\left[Z_{\nu}^n\right]}\right)^2\;,\\
G_2&=\frac{\mathbb{E}_{\phi}\left[\moy{z^3}_{\nu}\moy{z}_{\nu}Z_{\nu}^n\right]}{\mathbb{E}_{\phi}\left[Z_{\nu}^n\right]}+2r_0\frac{\mathbb{E}_{\phi}\left[\left(\moy{z^2}_{\nu}+\moy{z}_{\nu}^2\right)Z_{\nu}^n\right]}{\mathbb{E}_{\phi}\left[Z_{\nu}^n\right]}-\left(\frac{\mathbb{E}_{\phi}\left[\moy{z^2}_{\nu}Z_{\nu}^n\right]}{\mathbb{E}_{\phi}\left[Z_{\nu}^n\right]}\right)\left(\frac{\mathbb{E}_{\phi}\left[\moy{z}_{\nu}^2 Z_{\nu}^n\right]}{\mathbb{E}_{\phi}\left[Z_{\nu}^n\right]}\right)\;,\nn\\
G_3&=\frac{\mathbb{E}_{\phi}\left[\moy{z^2}_{\nu}^2 Z_{\nu}^n\right]}{\mathbb{E}_{\phi}\left[Z_{\nu}^n\right]}+4r_0\frac{\mathbb{E}_{\phi}\left[\moy{z}_{\nu}^2 Z_{\nu}^n\right]}{\mathbb{E}_{\phi}\left[Z_{\nu}^n\right]}-\left(\frac{\mathbb{E}_{\phi}\left[\moy{z^2}_{\nu}Z_{\nu}^n\right]}{\mathbb{E}_{\phi}\left[Z_{\nu}^n\right]}\right)^2\;,\\
G_4&=\frac{\mathbb{E}_{\phi}\left[\moy{z^2}_{\nu}^2 Z_{\nu}^n\right]}{\mathbb{E}_{\phi}\left[Z_{\nu}^n\right]}+2r_0\frac{\mathbb{E}_{\phi}\left[\left(\moy{z^2}_{\nu}+\moy{z}_{\nu}^2\right)Z_{\nu}^n\right]}{\mathbb{E}_{\phi}\left[Z_{\nu}^n\right]}-\left(\frac{\mathbb{E}_{\phi}\left[\moy{z}_{\nu}^2 Z_{\nu}^n\right]}{\mathbb{E}_{\phi}\left[Z_{\nu}^n\right]}\right)^2\;,\\
G_5&=\frac{\mathbb{E}_{\phi}\left[\moy{z^2}_{\nu}\moy{z}_{\nu}^2 Z_{\nu}^n\right]}{\mathbb{E}_{\phi}\left[Z_{\nu}^n\right]}+r_0\frac{\mathbb{E}_{\phi}\left[\left(\moy{z^2}_{\nu}+3\moy{z}_{\nu}^2\right)Z_{\nu}^n\right]}{\mathbb{E}_{\phi}\left[Z_{\nu}^n\right]}-\left(\frac{\mathbb{E}_{\phi}\left[\moy{z}_{\nu}^2 Z_{\nu}^n\right]}{\mathbb{E}_{\phi}\left[Z_{\nu}^n\right]}\right)^2\;,\\
G_6&=\frac{\mathbb{E}_{\phi}\left[\moy{z^2}_{\nu}\moy{z}_{\nu}^2 Z_{\nu}^n\right]}{\mathbb{E}_{\phi}\left[Z_{\nu}^n\right]}+4r_0\frac{\mathbb{E}_{\phi}\left[\moy{z}_{\nu}^2Z_{\nu}^n\right]}{\mathbb{E}_{\phi}\left[Z_{\nu}^n\right]}-\left(\frac{\mathbb{E}_{\phi}\left[\moy{z}_{\nu}^2 Z_{\nu}^n\right]}{\mathbb{E}_{\phi}\left[Z_{\nu}^n\right]}\right)\left(\frac{\mathbb{E}_{\phi}\left[\moy{z^2}_{\nu} Z_{\nu}^n\right]}{\mathbb{E}_{\phi}\left[Z_{\nu}^n\right]}\right)\\
G_7&=\frac{\mathbb{E}_{\phi}\left[\moy{z}_{\nu}^4 Z_{\nu}^n\right]}{\mathbb{E}_{\phi}\left[Z_{\nu}^n\right]}+4r_0\frac{\mathbb{E}_{\phi}\left[\moy{z}_{\nu}^2Z_{\nu}^n\right]}{\mathbb{E}_{\phi}\left[Z_{\nu}^n\right]}-\left(\frac{\mathbb{E}_{\phi}\left[\moy{z}_{\nu}^2 Z_{\nu}^n\right]}{\mathbb{E}_{\phi}\left[Z_{\nu}^n\right]}\right)^2\;.
\end{align}

In order to simplify these expressions, we can now use the following identity
\begin{align}
 &\partial_{x}\int dz\,z^m\,e^{-\frac{\beta\nu}{2}z^2-\beta\phi(z+x)}=\partial_{x}\int dz\,(z-x)^m\,e^{-\frac{\beta\nu}{2}(z-x)^2-\beta\phi(z)}\\
 &=\int dz\,(z-x)^{m-1}\left[\beta\nu(z-x)^2-m\right]\,e^{-\frac{\beta\nu}{2}(z-x)^2-\beta\phi(z)}=\int dz\,z^{m-1}\left[\beta\nu\,z^2-m\right]\,e^{-\frac{\beta\nu}{2}z^2-\beta\phi(z+x)}\;,\nn
\end{align}
which, exploiting the stationarity of the disordered potential $\phi(z)$ allows to derive the identity, valid for any integer $k$
\begin{align}
&\mathbb{E}_{\phi}\left[\partial_x\left(\prod_{j=1}^k\moy{z^{m_j}}_{\nu}Z_{\nu}^n\right)\right]_{x=0}=0 \label{id_moments}\\
&=(n-k)\beta\nu\mathbb{E}_{\phi}\left[\moy{z}_{\nu}\left(\prod_{j=1}^k\moy{z^{m_j}}_{\nu}Z_{\nu}^n\right)\right]+\sum_{i=1}^k\mathbb{E}_{\phi}\left[\left(\beta \nu \moy{z^{m_i+1}}_{\nu}-m_i\moy{z^{m_i-1}}_{\nu}\right)\prod_{j\neq i}\moy{z^{m_j}}_{\nu}Z_{\nu}^n\right]\;.\nn
\end{align}
In particular, one can derive the following identities by specializing in Eq. \eqref{id_moments} various values of parameters $k$ and $m_j$:
\begin{align}
&\frac{\mathbb{E}_{\phi}\left[\moy{z^2}_{\nu}Z_{\nu}^n\right]}{\mathbb{E}_{\phi}\left[Z_{\nu}^n\right]}+(n-1)\frac{\mathbb{E}_{\phi}\left[\moy{z}_{\nu}^2\,Z_{\nu}^n\right]}{\mathbb{E}_{\phi}\left[Z_{\nu}^n\right]}=\frac{1}{\beta\nu}\;,\label{id_app_1}\\
    &\frac{\mathbb{E}_{\phi}\left[\moy{z^4}_{\nu}Z_{\nu}^n\right]}{\mathbb{E}_{\phi}\left[Z_{\nu}^n\right]}+(n-1)\frac{\mathbb{E}_{\phi}\left[\moy{z^3}_{\nu}\moy{z}_{\nu}\,Z_{\nu}^n\right]}{\mathbb{E}_{\phi}\left[Z_{\nu}^n\right]}=\frac{3}{\beta\nu}\frac{\mathbb{E}_{\phi}\left[\moy{z^2}_{\nu}Z_{\nu}^n\right]}{\mathbb{E}_{\phi}\left[Z_{\nu}^n\right]}\;,\\
&\frac{\mathbb{E}_{\phi}\left[\moy{z^3}_{\nu}\moy{z}_{\nu}\,Z_{\nu}^n\right]}{\mathbb{E}_{\phi}\left[Z_{\nu}^n\right]}+\frac{\mathbb{E}_{\phi}\left[\moy{z^2}_{\nu}^2\,Z_{\nu}^n\right]}{\mathbb{E}_{\phi}\left[Z_{\nu}^n\right]}+(n-2)\frac{\mathbb{E}_{\phi}\left[\moy{z^2}_{\nu}\moy{z}_{\nu}^2\,Z_{\nu}^n\right]}{\mathbb{E}_{\phi}\left[Z_{\nu}^n\right]}=\frac{\mathbb{E}_{\phi}\left[\left(\moy{z^2}_{\nu}+2\moy{z}_{\nu}^2\right)Z_{\nu}^n\right]}{\beta\nu\mathbb{E}_{\phi}\left[Z_{\nu}^n\right]}\;,\\
&3\frac{\mathbb{E}_{\phi}\left[\moy{z^2}_{\nu}\moy{z}_{\nu}^2\,Z_{\nu}^n\right]}{\mathbb{E}_{\phi}\left[Z_{\nu}^n\right]}+(n-3)\frac{\mathbb{E}_{\phi}\left[\moy{z}_{\nu}^4\,Z_{\nu}^n\right]}{\mathbb{E}_{\phi}\left[Z_{\nu}^n\right]}=\frac{3}{\beta\nu}\frac{\mathbb{E}_{\phi}\left[\moy{z}_{\nu}^2 Z_{\nu}^n\right]}{\mathbb{E}_{\phi}\left[Z_{\nu}^n\right]}\;.
\end{align}

Using the expressions for the different coefficients and identities above, we are now in a position to compute explicitly the eigenvalues of the quadratic form. The first set of eigenvalues is obtained by summing Eq. \eqref{eq_eig_aa} over $a$ and Eq. \eqref{eq_eig_ab} over $a\neq b$ respectively, yielding
\begin{align}
    \Lambda_n\, \sum_{a}\eta_{aa}=&\left(G_{aaaa}+(n-1)G_{aabb}\right)\sum_{a}\eta_{aa}+\left(2G_{aaba}+(n-2)G_{aacd}\right)\sum_{a\neq b}\eta_{ab}\\
    \Lambda_n\, \sum_{a\neq b}\eta_{ab}=&(n-1)\left(2G_{aaab}+(n-2)G_{aabc}\right)\sum_{a}\eta_{aa}\\
    &+\left(G_{abba}+G_{abab}+2(n-2)(G_{abac}+G_{abca})+(n-2)(n-3)G_{abcd}\right)\sum_{a\neq b}\eta_{ab}\nn\;.
\end{align}
One can show that the system above can be put in the following form
\begin{align}
 \Lambda_n\begin{pmatrix}
    \sum_{a}\eta_{aa}\\
    \sum_{a\neq b}\eta_{ab}
\end{pmatrix}&=\begin{pmatrix}
    \rho_n+A_n & A_n\\
   (n-1)A_n & \rho_n+(n-1)A_n
\end{pmatrix}\begin{pmatrix}
    \sum_{a}\eta_{aa}\\
    \sum_{a\neq b}\eta_{ab}
\end{pmatrix}\;,
\end{align}
such that the two eigenvalues take the form
\be
\Lambda_n=\rho_n\;,\;\;\Lambda_n=\rho_n+n A_n\;.
\ee
We are interested in the limit $n=0$ where these eigenvalues are degenerate and read
\be
\Lambda_0=\rho_0=-\frac{1}{2(\beta\mu)^2}\;.\label{Lamb_0}
\ee

The second set of eigenvalues is obtained by ensuring that $\sum_a \eta_{aa}=0$ and $\sum_{a\neq b}\eta_{ab}=0$. One thus obtains from Eq. \eqref{eq_eig_aa} and Eq. \eqref{eq_eig_ab} respectively
\begin{align}
\Lambda_n\eta_{aa}=&\left(G_{aaaa}-G_{aabb}\right)\eta_{aa}+2\left(G_{aaba}-G_{aacd}\right)\sum_{b}\eta_{ab}(1-\delta_{ab})\;,\\
   \Lambda_n\sum_{b}\eta_{ab}(1-\delta_{ab})=&(n-2)\left(G_{aaab}-G_{aabc}\right)\eta_{aa}\\
&+\left(G_{abba}+G_{abab}+(n-4)\left(G_{abac}+G_{abbc}\right)-2(n-3)G_{abcd}\right)\sum_{b}\eta_{ab}(1-\delta_{ab})\nn\;.
\end{align}
One can show that the system above can be put in the following form
\begin{align}
 \Lambda_n\begin{pmatrix}
    \eta_{aa}\\
    \sum_{b}\eta_{ab}(1-\delta_{ab})
\end{pmatrix}&=\begin{pmatrix}
    \sigma_n+B_n & B_n\\
   \left(\frac{n}{2}-1\right)B_n & \sigma_n+\left(\frac{n}{2}-1\right)B_n
\end{pmatrix}\begin{pmatrix}
    \eta_{aa}\\
    \sum_{b}\eta_{ab}(1-\delta_{ab})
\end{pmatrix}\;,
\end{align}
such that the two eigenvalues take the form
\be
\Lambda_n=\sigma_n\;,\;\;\Lambda_n=\sigma_n+\frac{n}{2} B_n\;.
\ee
Taking again the limit $n=0$, these eigenvalues are degenerate and coincide with the eigenvalues previously obtained in Eq. \eqref{Lamb_0}.

The last eigenvalue is obtained by ensuring that both $\eta_{aa}=0$ and $\sum_{b}\eta_{ab}(1-\delta_{ab})=0$. It can be obtained directly from Eq. \eqref{eq_eig_ab} and reads
\begin{align}
\Lambda_n=G_{abba}+G_{abab}-2G_{abac}-2G_{abca}+2G_{abcd}\;.
\end{align}
In the limit $n\to 0$ and for $\beta\to \infty$, the rescaled eigenvalue reads
\begin{align}
\lim_{\beta\to\infty}2(\beta\mu)^2\Lambda_{n=0}=\lambda_{\rm RS}&=\alpha\mathbb{E}_{\phi}\left[\lim_{\beta\to \infty}\left(\beta\mu(\moy{z^2}_{\mu}-\moy{z}_{\mu}^2)\right)^2\right]-(1+\alpha)\;.\label{replicon_app_z}
\end{align}
Let us first re-express this eigenvalue explicitly in terms of the properties of the curvature  of the effective one-dimensional Hamiltonian Eq. \eqref{H_e_min}. The rescaled thermal variance of the position appears explicitly in the expression above. The latter can be obtained from the rescaled cumulant generating function
as 
\be
\mu\left.\partial_{h}^2\lim_{\beta\to \infty}\frac{1}{\beta}\ln \moy{e^{\beta h z}}_{\mu}\right|_{h=0}=\lim_{\beta\to \infty}\left(\beta\mu(\moy{z^2}_{\mu}-\moy{z}_{\mu}^2)\right)\;.
\ee
This rescaled cumulant generating function has been introduced in the main text and can be expressed in terms of the ground-state energy difference
\be
\lim_{\beta\to \infty}\frac{1}{\beta}\ln \moy{e^{\beta h z}}_{\nu}=\lim_{\beta\to \infty}\frac{1}{\beta}\ln \frac{\displaystyle\int e^{-\beta\left[\frac{\nu}{2}z^2-h z+\phi(z)\right]}\,dz}{\displaystyle\int e^{-\beta\left[\frac{\nu}{2}z^2+\phi(z)\right]}\,dz}=\epsilon_{\min}(\nu)-\epsilon_{\min}(\nu,h)\;,
\ee
where the ground-state energy of the one-dimensional model is defined as
\be
\epsilon_{\min}(\nu,h)=\min_z\left[\frac{\nu}{2}z^2-h z+\phi(z)\right]\;.
\ee
In particular from this expression, one can easily derive
\be
-\partial_h \epsilon_{\min}(\nu,h)=z_{\min}(\nu,h)\;,\;\; -\partial_h^2\epsilon_{\min}(\nu,h)=\partial_h z_{\min}(\nu,h)=\frac{1}{\nu+\phi''(z_{\min})}\;.
\ee
Finally, one can show that
\be
\lim_{\beta\to \infty}\left(\beta\mu(\moy{z^2}_{\mu}-\moy{z}_{\mu}^2)\right)=-\mu\partial_h^2\epsilon_{\min}(\mu,0)=\frac{\mu}{\mu+\phi''(z_{\min})}\;.
\ee
Inserting this identity in Eq. \eqref{replicon_app_z}, one obtains the final expression for the replicon eigenvalue in Eq. \eqref{replicon_eigv}. Note that from identity Eq. \eqref{id_app_1} obtained above, one can show
\be
\mathbb{E}_{\phi}\left[\lim_{\beta\to \infty}\left(\beta\mu(\moy{z^2}_{\mu}-\moy{z}_{\mu}^2)\right)^2\right]\geq\mathbb{E}_{\phi}\left[\lim_{\beta\to \infty}\left(\beta\mu(\moy{z^2}_{\mu}-\moy{z}_{\mu}^2)\right)\right]^2=\mathbb{E}_{\phi}\left[\frac{\mu}{\mu+\phi''(z_{\min})}\right]^2=1\;,
\ee
showing thus that the rescaled eigenvalue in Eq. \eqref{replicon_app_z} is indeed the largest  eigenvalue of the quadratic form.

\section{Matching the replica-symmetry breaking and the trivialisation transitions} \label{match_crit}

As demonstrated in this article, many properties of the $N$-dimensional problem may be expressed in terms of the one-dimensional disordered Hamiltonian 
\be
H_{\mu,h}(z)=\frac{\mu}{2}z^2-h z+\phi(z)\;,
\ee
and in particular its ground-state energy $\epsilon_{\min}(\mu,h)$ and position $z_{\min}(\mu,h)$ defined as
\begin{align}
    \epsilon_{\min}(\mu,h)&=\min_z H_{\mu,h}(z)\;,\\
    z_{\min}(\mu,h)&=\underset{z}{\rm argmin}\, H_{\mu,h}(z)\;.
\end{align}
To analyse the properties of the RS phase, it is sufficient to focus on the case where $h=0$. The number of minima $n_{\min}(\mu)$ of this one-dimensional disordered system is a random variable which must satisfy $n_{\min}(\mu)\geq 1$. Its mean can be computed explicitly using the simplest variant of the Kac-Rice formula:
\be
\Ep{n_{\min}(\mu)}=\int \Ep{\delta(\mu z+\phi'(z))(\mu+\phi''(z))\Theta(\mu+\phi''(z))\,dz}\;.
\ee
Denoting $G=\phi'(z)$ and $T=\phi''(z)$ the two random variables and recalling that their statistics is independent of $z$, one can simply check that
\begin{align}
  \Ep{n_{\min}(\mu)}&=\mathbb{E}\left[\int \delta(\mu z+G)(\mu+T)\Theta(\mu+T)\,dz\right]=\mathbb{E}\left[\frac{\mu+T}{\mu}\Theta(\mu+T)\right]\\
  &=\int_{-\mu}^{\infty} \frac{\mu+t}{\mu}p_0(t)\,dt=1+\mu\int_{1}^{\infty}\,(x-1)\,p_0(-\mu x)\,dx\ge 1\;,  
\end{align}
where we have denoted as in the main text $p_0(t)$ the PDF of the random variable $T=\phi''(z)$.  Supposing that the PDF $p_0(t)$ has a finite left edge $-t_{\rm e}$, it is not difficult to realise that $\Ep{n_{\min}(\mu)}=n_{\min}(\mu)=1$ for any $\mu>t_{\rm e}$. In that regime, there is a single minimum to the one-dimensional random landscape, which is necessarily the ground-state energy. The associated probability density of  $\chi_{\min}=\phi''[z_{\min}(\mu)]$ then reads
\be
p_*(t)=\frac{\mu+t}{\mu}p_0(t)\;,\;\;\mu>t_{\rm e}\;.
\ee
Using this result, one may in particular evaluate the replicon eigenvalue as 
\be
\lambda_{\rm RS}=\alpha\mathbb{E}_{\phi}\left[\frac{\mu^2}{\left(\mu+\phi''[z_{\min}(\mu)]\right)^{2}}-1\right]-1=\alpha\left(\int dt\,\frac{p_0(t)\,\mu}{\mu+t}-1\right)-1\;,\;\;\mu>t_{\rm e}\;.
\ee
We thus see that criterion for vanishing of the replicon eigenvalue perfectly coincides with that for the topology trivialisation transition. 

\section{An example of exactly solvable model}\label{ex_model}

Let us consider the one-dimensional Hamiltonian in Eq. \eqref{H_e_min} with the disordered potential
\be
\phi(z)=\gamma \cos(z-\theta)\;,
\ee
where $\theta$ is uniformly distributed in $[-\pi,\pi]$ while $\gamma$ is a positive random variable. For a fixed realisation of the random variable $\gamma$, this model can be rescaled to yield
\begin{align}
 H_{\nu,0}(z)&=\frac{\nu}{2}z^2+\gamma\cos(z-\theta)=\gamma\,\tilde H_{\nu/\gamma,0}(z)\;,\\
 \tilde H_{\nu,0}(z)&=\frac{\nu}{2}z^2+\cos(z-\theta)\;.
\end{align}
We denote $\tilde z_{\min}(\nu,\theta)$ the position of its minimum and $\tilde H_{\nu}(z,\theta)$ the value of its ground-state energy. The position $\tilde z_{\min}(\nu,\theta)$ of the global minimum $\tilde \epsilon_{\min}(\nu,\theta)$ lies in the interval $[-\pi,\pi]$ as
$\tilde H_{\nu}(z,\theta)-\tilde H_{\nu}(z+2\pi,\theta)=\nu/2\left[z^2-(z+2\pi)^2\right]\leq 0$. The random variable $\Delta=\tilde z_{\min}-\theta$ satisfies
\begin{align}
   &\nu(\Delta+\theta)-\sin(\Delta)=0\;,\label{z_min_eq}\\
&\nu-\cos(\Delta)\geq 0\;. 
\end{align}
Let us denote $y(\theta,\nu)$ the solution (in terms of $\Delta$) of Eq. \eqref{z_min_eq}. The PDF of the angle difference $\Delta$ can be obtained exactly as
\begin{align}
  {\cal P}_{\nu}(\Delta)&=\frac{\displaystyle\int  \Theta(\nu-\cos(y(\theta,\nu)))\,\delta\left[\Delta-y(\theta,\nu)\right]\,d\theta}{\displaystyle\int \Theta(\nu-\cos(y(\theta,\nu)))\,d\theta}=\frac{\nu-\cos(\Delta)}{2\pi\nu}\Theta(\Delta^2-g(\nu)^2)\;,  
\end{align}
where $g(\nu)=0$ if $\nu>1$ and $g(\nu)=y(0,\nu)$ otherwise. The distribution of all relevant random variables can be computed from that of $\Delta$, using in particular
\begin{align}
\tilde z_{\min}&=\frac{1}{\nu}\sin(\Delta)\;,\;\;\phi''(\tilde z_{\min})=-\cos(\Delta)\;,\\
\tilde \epsilon_{\min}&=\frac{1}{2\nu}\sin(\Delta)^2+\cos(\Delta)\;.   
\end{align}
The distribution of $\tilde z_{\min}$ reads in particular
\begin{align}
    p_{\nu}(z)=\begin{cases}
    \displaystyle\frac{1}{2\pi}\left[1+\frac{\nu}{\sqrt{1-(\nu z)^2}}\right]\Theta\left(\sin(g(\nu))^2-z^2\right)\;,\;\;0<\nu<\frac{2}{\pi}\;,\\
    \displaystyle\frac{1}{2\pi}\left[1+\frac{\nu}{\sqrt{1-(\nu z)^2}}\right]+\frac{1}{2\pi}\left[-1+\frac{\nu}{\sqrt{1-(\nu z)^2}}\right]\Theta\left(z^2-\sin(g(\nu))^2\right)\;,\;\;\frac{2}{\pi}<\nu<1\;,\\
    \displaystyle\frac{\nu}{\pi\sqrt{1-(\nu z)^2}}\;,\;\;\nu>1\;.
    \end{cases}
\end{align}
The distribution $p_*(t)$ of $\phi''(z_{\min})$ can be obtained explicitly as well and reads
\be
p_*(t)=\frac{\nu+t}{\pi\nu\sqrt{1-t^2}}\Theta(1-t)\Theta(t+t_m(\nu))\;,
\ee
where the value of $t_m(\nu)=\cos(g(\nu))$ is obtained by ensuring that this PDF is normalised. One can then simply check that the distribution of the Hessian $p_*(t-\nu)$ displays a finite gap $\nu-t_m(\nu)>0$ away from zero for any value of $\nu\neq 1$, while the PDF vanishes linearly as the Hessian goes to zero precisely for $\nu=1$.

Finally, some important quantities can be computed explicitly for this model
\begin{align}
\Ep{\left(\frac{\mu}{\mu+\phi''[z_{\min}]}\right)^m}&=\mathbb{E}_{\gamma}\left[\int d\nu\,\delta\left(\nu-\frac{\mu}{\gamma}\right)\int_{-\pi}^{\pi} d\Delta \,{\cal P}_{\nu}(\Delta) \,\left(\frac{\nu}{\nu-\cos(\Delta)}\right)^m\right]\nn\\
&=\mathbb{E}_{\gamma}\left[\int d\nu\,\delta\left(\nu-\frac{\mu}{\gamma}\right)\int_{g(\nu)}^{\pi} \frac{d\Delta}{\pi}\left(\frac{\nu}{\nu-\cos(\Delta)}\right)^{m-1}\right]\;.    
\end{align}
For $m=1$, it yields 
\be
\int_{-\pi}^{\pi}{\cal P}_{\nu}(\Delta) \,\left(\frac{\nu}{\nu-\cos(\Delta)}\right)\, d\Delta=1-\frac{g(\nu)}{\pi} \;,
\ee
while for $m=2$, 
\begin{align}
&\int_{-\pi}^{\pi} {\cal P}_{\nu}(\Delta) \,\left(\frac{\nu}{\nu-\cos(\Delta)}\right)^2 \,d\Delta=\int_{g(\nu)}^{\pi} \left(\frac{\nu}{\nu-\cos(\Delta)}\right)\,\frac{d\Delta}{\pi}\\
&=\begin{cases}
\displaystyle\frac{2\nu}{\sqrt{1-\nu^2}}\left[{\rm arctanh}\left(\frac{\sqrt{1-\nu^2}\cos(g(\nu))}{\cos(g(\nu))+\nu(\sqrt{1-\cos(g(\nu))^2}-1)}\right)+{\rm arctanh}\left(\sqrt{\frac{1-\nu}{1+\nu}}\right)\right]&\;,\;\;0<\nu<1\;,\\
  \displaystyle  \frac{\nu}{\sqrt{\nu^2-1}}&\;,\;\;\nu>1\;.
\end{cases}
\end{align}

Note that for a more general model considered in our paper 
\be
H_{\nu,h}(z)=\frac{\nu}{2}z^2- h z+\gamma \cos(z-\theta)=\gamma\left[\frac{\nu}{2\gamma}\left(z-\frac{h}{\nu}\right)^2+\cos\left(\left(z-\frac{h}{\nu}\right)-\left(\theta-\frac{h}{\nu}\right)\right)\right]-\frac{h^2}{2\nu}\;.
\ee
Using the stationarity of $\phi(z)$ for fixed $h$ and $\gamma$, the statistics of the global minimum $\epsilon_{\min}$ and its position $z_{\min}$ can simply be related to those of the  model studied above as
\be
\epsilon_{\min}\equiv\gamma\, \tilde \epsilon_{\min}\left(\frac{\nu}{\gamma}\right)-\frac{h^2}{2\nu}\;,\;\;z_{\min}\equiv\tilde z_{\min}\left(\frac{\nu}{\gamma}\right)-\frac{h}{\nu}\;.
\ee

\noindent{\it Phase diagram}

Let us briefly discuss the phase diagram for this example in the simplest example where $\gamma=1$ is fixed. In that case, the de-Almeida Thouless line is obtained as the solution of 
\begin{align}
\lambda_{\rm RS}=\alpha \left(\Ep{\left(\frac{\mu}{\mu+\phi''[z_{\min}]}\right)^2}-1\right)-1=\alpha \left(\frac{\mu}{\sqrt{\mu^2-1}}-1\right)-1=0\;,    
\end{align}
which yields
\be
\mu_c(\alpha)=\frac{1+\alpha}{\sqrt{1+2\alpha}}\;.
\ee
The pre-factors associated to the quadratic vanishing of the total complexity can be obtained explicitly as
\be
\partial_\mu^2 \Sigma_{\rm tot}(\mu_c(\alpha),\alpha)=\frac{2\alpha^2(1+2\alpha)^2}{(1+\alpha)^2(5+\alpha(22+\alpha(25+4\alpha)))}
\ee
and similarly for the complexity of minima, one finds 
\be
\partial_\mu^2 \Sigma_{\min}(\mu_c(\alpha),\alpha)=\frac{(1+2\alpha)^{5/2}}{(1+\alpha)(1+2\sqrt{1+2\alpha}+\alpha(2+\alpha+(6+2\alpha)\sqrt{1+2\alpha}))}\;.
\ee
One can similarly compute properties of the RSB phase, obtaining the rescaled breaking point as
\be
w_{\rm AT}=\frac{\alpha\mu^3\Ep{{\cal C}_3(\mu)^2}}{2\left(\alpha\mu^3\Ep{{\cal C}_2(\mu)^3}-(\alpha+2)\right)}=\frac{(1+\alpha)^2(1+2\alpha)^{3/2}(5+2\alpha(5+2\alpha))}{16 \alpha^4(1+\alpha(3+\alpha))}\;.
\ee
Finally, the derivative of the solution reads
\begin{align}
    w_{\rm AT}'=&\frac{\alpha\mu^4\left(\mathbb{E}\left[{\cal C}_4(\mu)^2\right]-12 w_{\rm AT}\,\mathbb{E}\left[{\cal C}_3(\mu)^2{\cal C}_2(\mu)\right]+6w_{\rm AT}^2\,\mathbb{E}\left[{\cal C}_2(\mu)^4\right]\right)-6(\alpha+3)w_{\rm AT}^2}{2\left(\alpha\mu^3\,\mathbb{E}\left[{\cal C}_2(\mu)^3\right]-(\alpha+2)\right)}\nn\\
    =&\frac{(\alpha+1)^3 (2 \alpha+1)^2}{512 \alpha^{10} (\alpha (\alpha+3)+1)^3}(1695+\alpha (20340+\alpha (102054+\alpha (278520+\alpha (451025\\
    &+16 \alpha (27760+\alpha (16401+\alpha (5507-4 \alpha
   (-228+(-10+\alpha) \alpha)))))))))\;.\nn
\end{align}
Interestingly this quantity changes from positive for $\alpha<\alpha_{\rm G}$ to negative for $\alpha>\alpha_{\rm G}$ with $\alpha_{\rm G}\approx22.9289..$

\section{Identities in the FRSB phase}\label{id_FRSB}

In this appendix, we provide some details on the derivation of identities in the FRSB phase that allow to derive the explicit expressions for the replicon eigenvalue in Eq. \eqref{replicon2}, the rescaled breaking point in Eq. \eqref{breaking_point2} and the slope of the Parisi function in Eq. \eqref{slope} at the breaking point. We remind that the functions $ P(t,h)$ and $f(t,h)$ satisfy the following equations
\begin{align}
    \partial_t P&=\frac{1}{2}\partial_{h}\left[\partial_{h}-2w(t)(\partial_{h}f)\right] P\;,\\
    \partial_t f&=-\frac{1}{2}\left[w(t)(\partial_{h}  f)^2+\partial_{h}^2 f\right]\;.
\end{align}
To simplify the expressions, we use the following rule
\be
\int r(h)\partial_t P\,dh=\frac{1}{2}\int r(h)\left[\partial_h^2 P-2w(l)\partial_h(P\partial_h f )\right]\,dh =\frac{1}{2}\int P\left(\partial_h^2+2w(l)\partial_h f\partial_h\right)r(h)\,dh\;.
\ee
Using this rule, one may simplify expressions as follows
\begin{align}
    &\int \partial_{t}\left(P  f\right)=\frac{1}{2}\int  P\left[\left(\partial_h^2  f+2w(t)(\partial_h  f)^2\right)-\left(\partial_h^2  f+w(t)(\partial_h f)^2\right)\right]\, dh\nn\\
    &=\frac{w(t)}{2}\int P(\partial_h f)^2\,dh\;,\\
    &\int \partial_{t}\left(P\left[\partial_{h}^2 f+w(t)(\partial_{h}  f)^2\right]\right)\,dh=\frac{1}{2}\int P\left(\partial_h^2 +2w(t)\partial_h f \partial_h\right)\left(\partial_{h}^2 f+w(t)(\partial_{h} f)^2\right)\,dh\nn\\
    &+w'(t)\int P(\partial_{h} f)^2\,dh-\frac{1}{2}\int P \left(\partial_h^2+2w(t)(\partial_h f)\partial_h\right)\left(\partial_h^2 f+w(t)(\partial_h f)^2\right)\,dh\nn\\
    &=w'(t)\int P(\partial_{h} f)^2\,dh\\
    &\int \partial_{t}\left(P (\partial_h  f)^2\right)\,dh=\frac{1}{2}\int P\left[\left(\partial_h^2 +2w(t)(\partial_h f)\partial_h\right)(\partial_h f)^2-2\partial_h f\left(\partial_h^2 f+w(t)(\partial_h f)^2\right)\right]\,dh\nn\\
    &=-\int P\left(\partial_h^2 f\right)^2\,dh\\
&\int \partial_t\left(P\left(\partial_{h}^2 f\right)^2\right)\,dh=\frac{1}{2}\int P\left[\left(\partial_h^2+2w(t)\partial_h f \partial_h\right)(\partial_{h} f)^2-2\partial_{h}^2 f\partial_h^2\left(\partial_{h}^2 f+w(t)(\partial_{h} f)^2\right)\right]\,dh\nn\\
&=-\int P\left[2w(t)\left(\partial_h^2 f\right)^3-\left(\partial_h^3 f\right)^2\right]\,dh\\
&\int \partial_t\left(P\left[2w(t)\left(\partial_h^2 f\right)^3-\left(\partial_h^3  f\right)^2\right]\right)\,dh=2w'(t)\int P\left(\partial_h^2 f\right)^3\,dh\nn\\
&+\frac{1}{2}\int  P\left(\partial_h^2+2w(t)\partial_h f \partial_h\right)\left[2w(t)\left(\partial_h^2 f\right)^3-\left(\partial_h^3 f\right)^2\right]\,dh\nn\\
&-\frac{1}{2}\int P\left[6w(t)\left(\partial_h^2 f\right)^2\partial_h^2-2\partial_h^3 f\partial_h^3\right]\left(\partial_h^2 f+2w(t)(\partial_h f)^2 \right)\,dh\nn\\
&=-\int P\left[6w(t)^2\left(\partial_h^2  f\right)^4-12w(t) \partial_h^2 f\left(\partial_h^3 f\right)^2+\left(\partial_h^4 f\right)^2\right]\,dh+2w'(t)\int P\left(\partial_h^2 f\right)^3\,dh\;.
\end{align}

\newpage
\bibliography{many_cosine_bibli}

\end{document}